\documentclass[aps,prl,reprint,showpacs,amssymb,floatfix]{revtex4-1}   
\usepackage{amsmath}    %
\usepackage{amsfonts}    %
\usepackage{graphicx}   %
\usepackage[caption=false]{subfig}  %
\usepackage{tikz}

\usepackage{xr}
\newcommand{\ket}[1]{\vert #1 \rangle}

\begin{document}
\title{Origin of Spinel Nanocheckerboards via First Principles}
\author{Mordechai Kornbluth}
\affiliation{Department of Applied Physics and Applied Mathematics, Columbia
 University, New York, NY 10027}
\author{Chris A. Marianetti}
\email{chris.marianetti@columbia.edu}
\affiliation{Department of Applied Physics and Applied Mathematics, Columbia
 University, New York, NY 10027}

\begin{abstract}
\noindent
Self-organizing nanocheckerboards have been experimentally fabricated in
 Mn-based spinels, but have not yet been explained with first principles.
Using density-functional-theory, we explain the phase diagram of the
 $\mathrm{ZnMn_xGa_{2-x}O_4}$ system and the origin of nanocheckerboards.
We predict total phase separation at zero temperature, then show the
combination of kinetics, thermodynamics, and Jahn-Teller physics that generates
 the system's observed behavior.
We find the \{011\} surfaces are strongly-preferred energetically, which
 mandates checkerboard ordering by purely geometrical considerations.
\end{abstract}

\date{\today}
\maketitle

Experimental observation demonstrates intriguing nanoscale compositional
 ordering in a variety of material alloys.
These include noble-metal-alloy nanocheckerboards
 \cite{Leroux1991PMB,Udoh1995MSEA,Winn2000JAC},
$\mathrm{BaTiO_3}$-$\mathrm{CoFe_2O_4}$ nanopillars \cite{Zheng2004Science},
and an assortment of manganite-spinel nanocheckerboards
\cite{Yeo2006APL,Park2008NL,OMalley2008PRB,
 Zhang2007APL133123,Zhang2007APL233110}.
Nanoscale phenomena are inherently difficult to treat with quantum mechanics'
first principles, due to the prohibitive scaling of electronic-structure
 methods.
Previous theoretical studies
 \cite{Bouar1998AM,Ni2007AM,Ni2008AM,Ni2009NL,Ni2009NM} %
used phase-field models \cite{Boettinger2002ARMR,Steinbach2009MSMSE} 
to simulate nanocheckerboard formation.  
However, those models rely upon coefficients chosen without first-principles
 justification.
In contrast, our work reveals the origin of nanocheckerboards from
 first-principles.

Here we examine the experimentally well-characterized manganite spinels
 $\mathrm{A^{2+}Mn_2^{3+}O_4^{2-}}$.
These cooperative-Jahn-Teller crystals, upon doping with certain transition
 metals, organize into nanocheckerboards.
Experiments showed that high-temperature mixing, followed by slow cooling,
yields a spontaneously-formed checkerboard whose squares alternate tetragonal
 Mn-rich and cubic Mn-poor phases.
These checkerboards emerge from the cross-section of spontaneously-aligned
 nanorods.
Yeo et al. \cite{Yeo2006APL} fabricated self-assembling nanocheckerboards from
$\mathrm{ZnGa_2O_4}$ (ZGO) + $\mathrm{ZnMn_2O_4}$ (ZMO), comprised of $\sim 4
 \times 4 \times 70 \textrm{ nm}^3$ nanorods.
Later work grew checkerboards with nanorods over 700 nm long on a MgO substrate
 \cite{OMalley2008PRB,Park2008NL}.
Checkerboards were later extended to other Mn-based spinels, 
first $\mathrm{MgMn_xFe_{2-x}O_4}$ (MMFO) \cite{Zhang2007APL133123}
and then the tunable-sized checkerboards of
 $\mathrm{Co_{0.6}Mn_xFe_{2.4-x}O_4}$ (CMFO) \cite{Zhang2007APL233110}.
The latter are notable for a patterning that alternates ferro- and paramagnetic
 phases,
which yields potential for ultrahigh-density information storage.
{
\begin{figure}[tbh] 
  \includegraphics[width=0.47\textwidth]{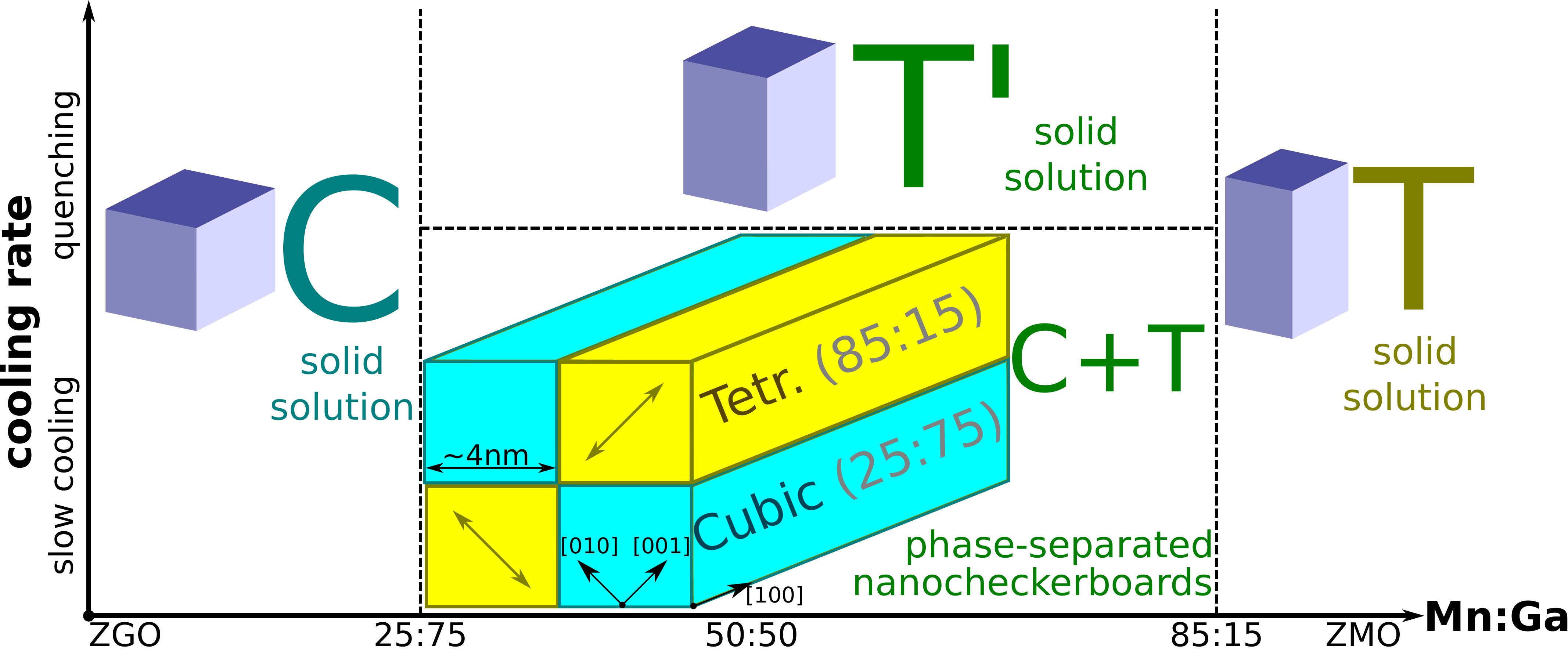}
\caption{Sketch of experimental results presented in Ref.
 \onlinecite{Yeo2006APL}.
For $x_{\mathrm{Mn}}\le 0.25$ a single-phase cubic structure (C) appears, while
 for $x_{\mathrm{Mn}}\ge 0.85$ it is a single-phase tetragonal structure (T).
For intermediate concentrations, a single-phase tetragonal structure (T$'$) is
observed upon rapid quenching (the high-temperature phase), while slower
cooling generates phase-separated nanocheckerboards with $(011)$ and
 $(01\overline{1})$ interfaces.
  Arrows in the tetragonal regions show the tetragonally-elongated direction. 
The concentrations on the abscissa are those for which data were reported in
 Ref. \onlinecite{Yeo2006APL}.
The concentrations within the nanocheckerboards were not reported as being
 measured directly.
  }
  \label{fig:experimental}
\end{figure}
}

We analyze the nanocheckerboard system $\mathrm{ZnMn_xGa_{2-x}O_4}$, whose
 relative simplicity renders it a minimal prototype.
We sketch the main experimental results of Ref. \onlinecite{Yeo2006APL} in
 Figure \ref{fig:experimental}.
ZGO is a cubic spinel while ZMO is a tetragonal spinel ($c/a=1.14$
 \cite{Asbrink1999PRB}), which
immediately leads to some anomalous differences between experiment and naive
 expectations.
First, room-temperature experiments reveal solid solutions at non-negligible
 concentrations,
despite the disparate crystal structures of the end-members,
in violation of the Hume-Rothery rules.
Second, x-ray diffraction experiments show a cubic structure up to 25\% ZMO
 (and 75\% ZGO),
whereas a non-negligible tetragonality would  be expected at this concentration,
regardless of the origin of the observed solubility.
Third, in the region where ZMGO phase-separates, experiment shows checkerboard
 formations instead of traditional spinodal decomposition.
To address these issues, we compute the energetics of
$\mathrm{ZnMn_xGa_{2-x}O_4}$ using density functional theory as implemented in
the Vienna Ab-initio Simulation Package (VASP)
 \cite{Kresse1993PRB,Kresse1994PRB,Kresse1996CMS,Kresse1996PRB}.
We use a plane-wave cutoff energy of 415 eV, the PW91 generalized gradient
approximation (GGA) functional \cite{Perdew1992PRB,Perdew1993PRB}, and
 projector augmented wave (PAW) based pseudopotentials \cite{Kresse1999PRB}.
All calculations are initialized with ferromagnetic ordering for simplicity,
a well-justified approach for the N\'{e}el temperature of merely $\approx 20$ K
 \cite{Asbrink1999PRB,Choi2006PRB,Li2011MCP}.
We compute a phase diagram via a cluster expansion \cite{Ruban2008RPP} of the
DFT energetics with Monte-Carlo simulations, as implemented in the Alloy
Theoretic Automated Toolkit (ATAT) package
 \cite{Walle2002Calphad,Walle2002JPE,Walle2002MSMSE,Walle2013JOM}.
Both ZGO and ZMO crystallize in the spinel structure, with nominal valences of
 $\mathrm{Zn^{2+}(Ga,Mn)_2^{3+}O_4^{2-}}$.
Zn occupies the tetrahedral (``A'') sites and Ga / Mn occupy the octahedral
(``B'') sites, with negligible inversion
 \cite{Fritsch2000SSI,Errandonea2009PRB}.
Therefore, our study focuses on the effect of Ga / Mn occupation of the B sites.
Each B-centered octahedron shares edges with six others, which couples their
 anionic distortions.
ZGO forms a cubic spinel (space group $Fd\bar{3}m$) \cite{Errandonea2009PRB},
while ZMO is a tetragonal spinel (space group $I4_1/amd$) with significant
 distortion $c/a=1.14$ \cite{Asbrink1999PRB}.
The T$'$ of Figure \ref{fig:experimental} (high-temperature fully-mixed ZMGO)
 has $c/a \approx 1.06$ \cite{Yeo2006APL}.
We have provided lattice constants from the literature, and their
 DFT-calculated analogs, in Supplementary Material (Table S1);
DFT calculations agree with experiment.
ZMO's sizable tetragonal distortion is due to the Jahn-Teller (JT) effect in
 the $\mathrm{Mn}^{3+}$ ions,
where the $d^4$ configuration in a high-spin octahedral environment causes the
 $e_g$ orbitals to break symmetry by a tetragonal distortion.
This orbital ordering leads to a martensitic cubic-to-tetragonal transition in
a variety of crystals, including spinels \cite{Kanamori1960JAP,Irani1962JPCS,
 Englman1972The,Gehring1975RPP,Kugel1982SPU}.
For consistency, we take [001] to be the JT-distorted direction.

Experiments, summarized in Figure \ref{fig:experimental}, show solid solutions
 for $x_{Mn}\le 0.25$ and $x_{Mn}\ge 0.85$.
Yet the cubic and strongly-tetragonal crystal structures of the end members are
 expected to be immiscible:
Placing a non-JT octahedron in a tetragonal JT-active environment, and vice
 versa, costs energy.
Our DFT calculations quantitatively verify the qualitative Hume-Rothery rules: 
We generated 192 supercells ($\le 42$ atoms) then fully relaxed the structures.
(See Supplementary Material for further calculation details.)
Figure \ref{fig:phases}(a) shows their formation energies, demonstrating that
 mixing the disparate crystal structures incurs an energy cost
\footnote{We would rather not blunt Occam's razor and trample intuition by
 suggesting a long-range-only phase-mixing effect with absolutely no evidence.}.
A ground-state search using the cluster expansion (via the aforementioned ATAT
 package) revealed no lower-energy supercells.
Therefore, the zero-temperature ground state of ZMGO is actually phase
separation into bulk ZGO and ZMO, a conclusion absent from previous
 non-first-principles analyses \cite{Ni2007AM,Ni2008AM,Ni2009NL,Ni2009NM}.
There is no chemical or physical reason to believe that ZGO and ZMO should mix
 at anything but elevated temperatures.

{
\begin{figure}[tbh] 
  \includegraphics[width=0.47\textwidth]{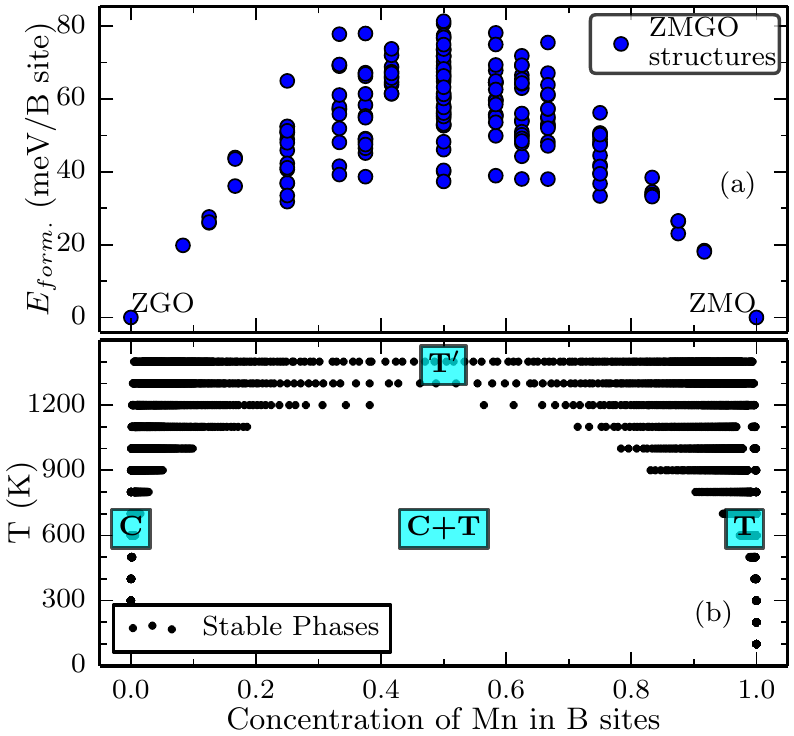}
\caption{ (a) Energies of formation of 192 $\mathrm{ZnMn_xGa_{2-x}O_4}$
 supercells.
  (b) Phase diagram of mixed ZGO and ZMO, using ATAT's EMC2 code.
The stable phases (with respect to a semi-grand-canonical ensemble, where
 $x_{\mathrm{Mn}}$ may vary) are shown at varying temperatures.
  The phases are marked:
Cubic (C), tetragonal (T), phase-separated (C+T), and high-temperature
 fully-mixed tetragonal ($\mathrm{T'}$).
  }
  \label{fig:phases}
\end{figure}
}

To further verify this, we computed a phase diagram via a cluster expansion,
 using the aforementioned ATAT package.
(See Supplementary Material for extensive calculation details.)
As shown in Figure \ref{fig:phases}(b), the zero-temperature stable phases are
 immiscible bulk ZGO and ZMO.
In fact, 
the observed miscibility at $x_{\mathrm{Mn}}\le 0.25$ and $x_{\mathrm{Mn}}\ge
 0.85$ is unreasonable at all but extreme temperatures.
We therefore attribute the anomalous miscibility to kinetic limitations, i.e.
 diffusion constraints.
This is corroborated by the experimental observation that slower cooling (i.e.
better diffusion) leads to larger checkers (i.e. less miscibility)
 \cite{Zhang2007APL133123,Zhang2007APL233110}.
In fact, kinetics are the obvious origin of the miscible $\mathrm{T}'$ state
 observed upon rapid quenching (shown in Figure \ref{fig:experimental}).
DFT's prediction for the $\mathrm{T}'$ state for $x_{\mathrm{Mn}}=0.5$ (taken
as the mean of all calculated structures) agrees well with the experimental
 measurements.
For example, DFT predicts $c/a=1.07$, comparable with experiment's $c/a \approx
 1.06$.
(See Supplementary Material for all data.)
Similarly, kinetic limitations must cause the apparent solubility of
 $x_{\mathrm{Mn}}\le 0.25$ and $x_{\mathrm{Mn}}\ge 0.85$.
For example, if diffusion essentially freezes by e.g. 900 K, the system cannot
separate into 100/0\% Mn mixtures and will remain a high-entropy frozen solid
 solution.
Unfortunately, quantitatively predicting kinetics (including checker size)
requires detailed knowledge of the diffusion mechanisms, coupled with complex
 JT lattice dynamics, which is beyond the scope of this Letter.
Having explained that the apparent solubility is due to kinetics,
we must address 
the second anomaly,
the experimental observation of a cubic structure for $x_{\mathrm{Mn}} \le
 0.25$.
In contrast,
DFT predicts a tetragonal structure ($c/a \approx 1.03$ for all calculated
structures at $x_{\mathrm{Mn}} = 0.25$) due to the cooperative Jahn-Teller
 effect.
We propose that the cubic structure of the Mn-poor phase is caused by
 noncooperative JT distortions at finite temperatures.
It is well-known that, as in many spinels, ZMO undergoes a phase transition to
a cubic spinel above approximately 1323 K, due to noncooperative JT distortions
(each octahedron distorting in a random direction)
 \cite{Irani1962JPCS,Fritsch2000SSI,Kanamori1960JAP}.
The transition temperature scales with doping:
Within mean-field theory it is approximately linear with doping
 \cite{Wojtowicz1959PR};
experimentally, a variety of spinels transition at $T_c \approx 9260 (c/a-1)$ K
 \cite{Irani1962JPCS},
where $c/a$ is the tetragonal distortion induced by the cooperative distortion
\footnote{Although our calculations show zero-temperature distortion, while the
empirical fitting is for room-temperature distortion, the same experiments show
relatively little change in distortion below the transition temperature, so the
 equation is still valid.}.
As illustrated in Figure \ref{fig:cubic_mnpoor}, both these trends imply that
at room temperature, structures of $x\lesssim 0.25$ are cubic due to this
 transition, while $x \gtrsim 0.25$ are tetragonal.
Similar ``critical concentrations'' of JT-ions have been documented in the
 literature \cite{Wojtowicz1959PR,Englman1970PRB}.
This is also consistent with the conclusion of Noh et al. \cite{Noh2006APL}
that although $x_{\mathrm{Mn}}=0.25$ has an XRD pattern of a cubic spinel, it
has a rather large JT splitting of $\sim 0.7$ eV, i.e. a high-spin electronic
 configuration, which leads to JT distortion.
(See Supplementary Material for detailed analysis of the reliability of GGA's
 high-spin prediction.)
Therefore, low-temperature measurements should reveal DFT's
 tetragonally-distorted structure for $x \le 0.25$.

{
\begin{figure}[tbh]
  \includegraphics[width=0.47\textwidth]{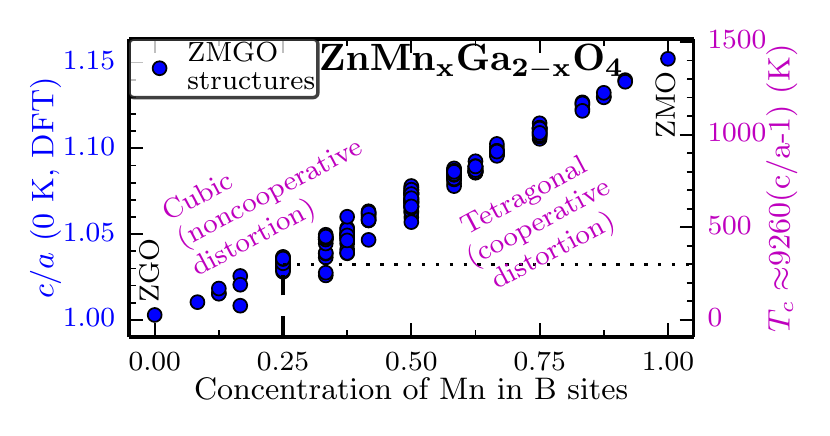}
  \caption{
  Phase diagram for the noncooperative Jahn-Teller transition in spinels.
DFT calculates the distortion (left axis) of our ZMGO structures (as in Figure
 \ref{fig:phases}(a)) at 0 K.
The critical temperature for the tetragonal-to-cubic transition (right axis) is
obtained based on the fitting $T_c \approx 9260 (c/a-1)$ K, empirically
 accurate for a variety of spinels \cite{Irani1962JPCS}.
  Therefore at room temperature, $x_{\mathrm{Mn}} \lesssim 0.25$ is cubic.
  }
  \label{fig:cubic_mnpoor}
\end{figure}
}

Therefore, the cubic structure is due to the high-entropy noncooperative
 distortion.
We should note that these finite-temperature Jahn-Teller lattice dynamics are
expected to enhance miscibility, as noted by other experiments
 \cite{Yeo2009JPCM}.
For example, whereas experiment places the boundaries at 25/85\% Mn, our phase
diagram (Figure \ref{fig:phases}(b)) shows that ZMO is more tolerant of
 Ga-doping than ZGO of Mn-doping.
This discrepancy is likely due to the lack of noncooperative distortions in the
cluster-expansion model, although a quantitative treatment is beyond the scope
 of this Letter.
However, this alone would not suffice to cause appreciable solubility near room
 temperature.
Now we turn to $0.25<x_{\mathrm{Mn}}<0.85$, 
where slowly-cooled samples phase-separate on a diffusion-limited scale.
We seek to explain how this leads to nanocheckerboards rather than traditional
 spinodal decomposition.
We calculate the energy of joining a slab of ZGO to a slab of ZMO along a
particular surface; lower energies indicate stable interface directions
 \cite{Liu2007PRL}.
Figure \ref{fig:slab_energies}(a) shows the energies of formation for various
 slab thicknesses.
(Larger thicknesses are inaccessible due to large, convergence-challenged
 supercells.)
In agreement with experiment, DFT prefers the \{011\} surfaces.
The energy relative to the next-preferred surface is about $10 \textrm{ meV/B}$
 (160 meV per cubic unit cell) at only 1.5 nm,
and presumably larger for the 4 nm nanocheckerboards.

{
\begin{figure}[tbh]
  \centering
  {
  \begin{tikzpicture} [
    ]
    \node[anchor=south] (a) at (0,0.0)
    { \includegraphics[width=0.47\textwidth]{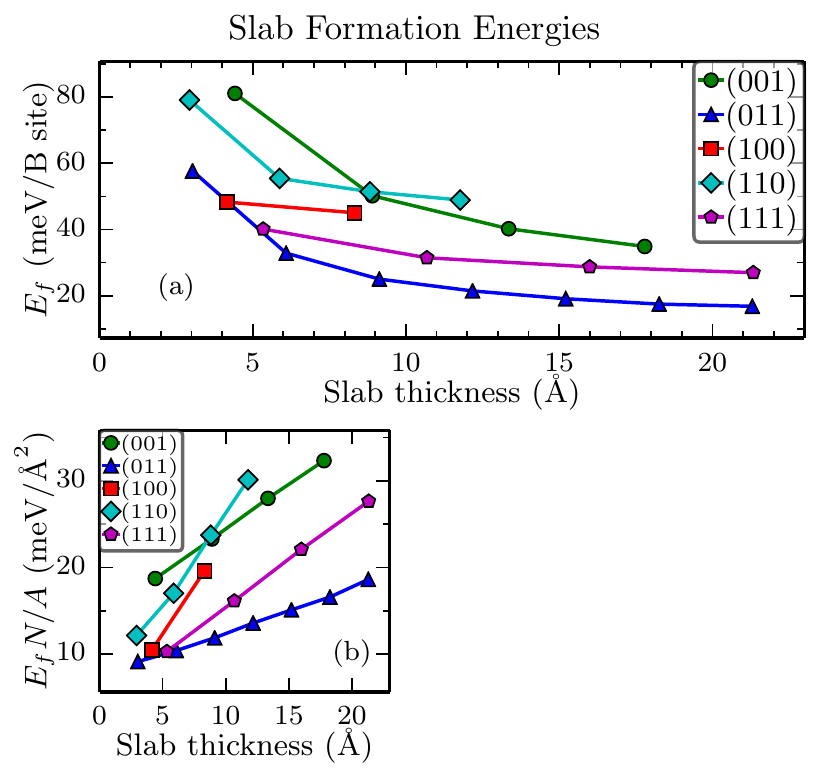}  };
    \node[anchor=south west] (b) at (0,0.3)
    {\includegraphics[width=0.22\textwidth]{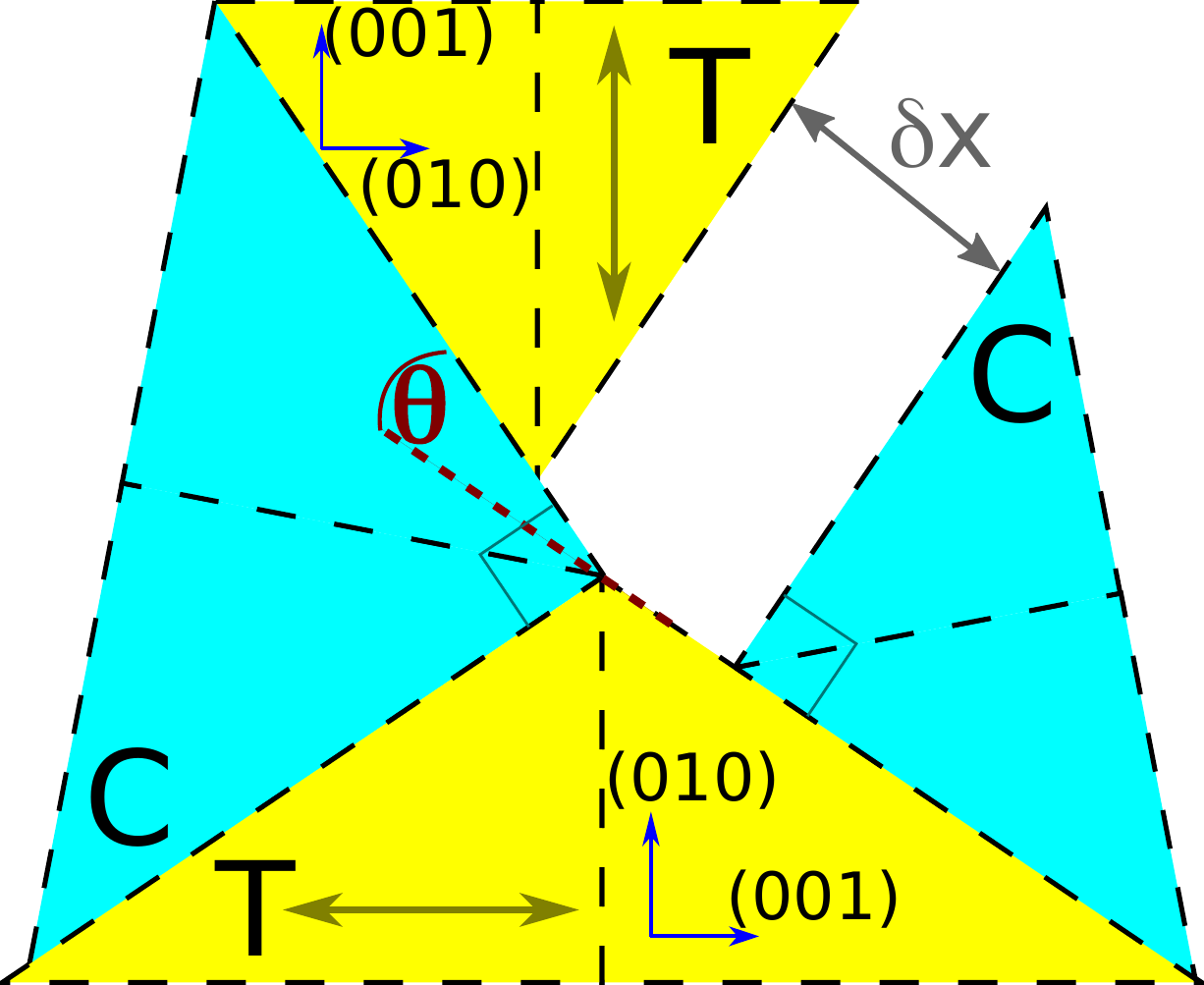}  };
    \node[anchor=north east] (c) at +(0.23\textwidth,3.7) {(c)};
  \end{tikzpicture}
  }
  \caption{
(a) Formation energies of ZGO/ZMO slabs layered in the given direction,
normalized per B-site, plotted against the average thickness of the ZGO and ZMO
 slabs.
(b) Same, multiplied by the number of B-sites ($N$) and divided by the
 interface area ($A$), giving a per-area normalization.
  (c)   Fitting cubic and tetragonal domains with a coherent interface.
  Arrows show the tetragonally elongated direction.
The \{011\} surfaces of [001]-distorted tetragonal  and cubic crystals form a
coherent interface within the lattice. Due to the bent angles at the
interfaces, this can occur only as checkerboards ($\delta x=0$, with a
 four-corner geometry). %
  The rotation $\theta$ can be found analytically (see text).
Dashed lines represent lattice constants, which must match for coherent
 interfaces.
  }
  \label{fig:slab_energies}
\end{figure}
}

Due to symmetry, only five directions are calculated directly.
We search formation energies of multilayer slabs of thickness $t$ oriented in
an arbitrary direction $\vec{\mathbf{k}}$ by performing an expansion in the
symmetry-adapted (tetragonal) harmonics, a standard group-theory methodology
 \cite{Laks1992PRB,Walle2014PRB}.
It is apparent from Figure \ref{fig:slab_energies}(b) that the total formation
 energy can be approximated by
$E(\vec{\mathbf{k}},t) = A( c_a( \vec{\mathbf{k}}) + t c_v (\vec{\mathbf{k}}))$
, where A is the cross-sectional area and $c_a,c_v$ are constants.
We perform an expansion of $c_a,c_v$ in the four lowest-order harmonics via a
 least-squares fit for the five calculated values of $\vec{\mathbf{k}}$.
(Fitting data appear in Supplementary Material.)
We thus confirm that (011), and equivalently $(01\overline{1})$, are the
 lowest-energy surfaces.

Hence physics dictates that when our system coherently mixes cubic and
 tetragonal phases, it forms $(011)$ and $(01\overline{1})$ interfaces.
Now pure geometry dictates checkerboard configurations for coherent interfaces.
As shown in Figure \ref{fig:slab_energies}(c), due to bent angles at the
interfaces, the only configuration that retains a coherent lattice is the
 checkerboard formation.
It was previously observed that alternating cubic phases are rotated by a few
degrees, while alternating tetragonal phases are rotated by $90^\circ$
 \cite{Yeo2006APL,Park2008NL,Zhang2007APL233110}.
The origin is now obvious from Figure \ref{fig:slab_energies}(c), with the
 cubic-phase angle of rotation
$\theta_c=\pi/2-2\tan^{-1}a/c $.
This agrees with measured values within $<1^\circ$ (data in Supplementary
 Material, Table S2).
This analysis relates our checkerboards to the CoPt cubic-tetragonal
nanostructures \cite{Leroux1991PMB,Bouar1998AM} and other lattice-induced
 interface rotations \cite{Honig2013NM}.

{
\iffalse %
%
%
%

The origin of the \{011\} preference is fascinating but beyond the scope of
 this work.
Previous work showed that both strain and anionic distortions
($\vec{\mathbf{k}}=0$ phonons) contribute significantly to the Jahn-Teller
 effect \cite{Marianetti2001PRB}.
By extracting the strain component from the slab formation energy, it appears
that here too, both are significant; calculation details appear in
 Supplementary Material.
In particular, the \{011\} interface contains favorable coupling between the
 strain mode and long-range (JT) anionic distortions.
In other words, long-range anionic distortions alleviate the energy penalties
 caused by lattice strain.
%

\fi
}{
\iftrue %

Whence does the \{011\} preference originate?
Previous work showed the importance of both strain and local ionic distortions
(i.e. $\vec{\mathbf{k}}=0$ phonons) to the Jahn-Teller effect
 \cite{Marianetti2001PRB}.
Our slabs' formation energy consists of 
strain energy, due to biaxial lattice-matching,
and contact energy, due to local ionic distortions and local binding energies. 
Specifically, for two slabs of thickness $t$ joined in direction
 $\hat{\mathbf{k}}$, we write:
\begin{equation}
E_{\textrm{formation}}=N E_{\textrm{strain}}(\hat{\mathbf{k}},t) + A
 E_{\textrm{contact}}(\hat{\mathbf{k}},t)
\end{equation}
where the units now ensure that the latter two $E$ terms converge for for $t
 \to \infty$.
As shown in Figure \ref{fig:decomposition_energies}(a), the strain energy can
be computed directly by relaxing ionic coordinates for ZGO and ZMO separately,
 with strained lattice vectors (the ``strained bulk'' calculation).
Then the contact energy is simply the difference between the formation energy
 and the strain energy.
(Detailed equations appear in Supplementary Material.)

{
\begin{figure}[tbh]
  \begin{tikzpicture}
\node[anchor=south east] (a) at (0,0.0) {
 \includegraphics[width=0.40\textwidth]{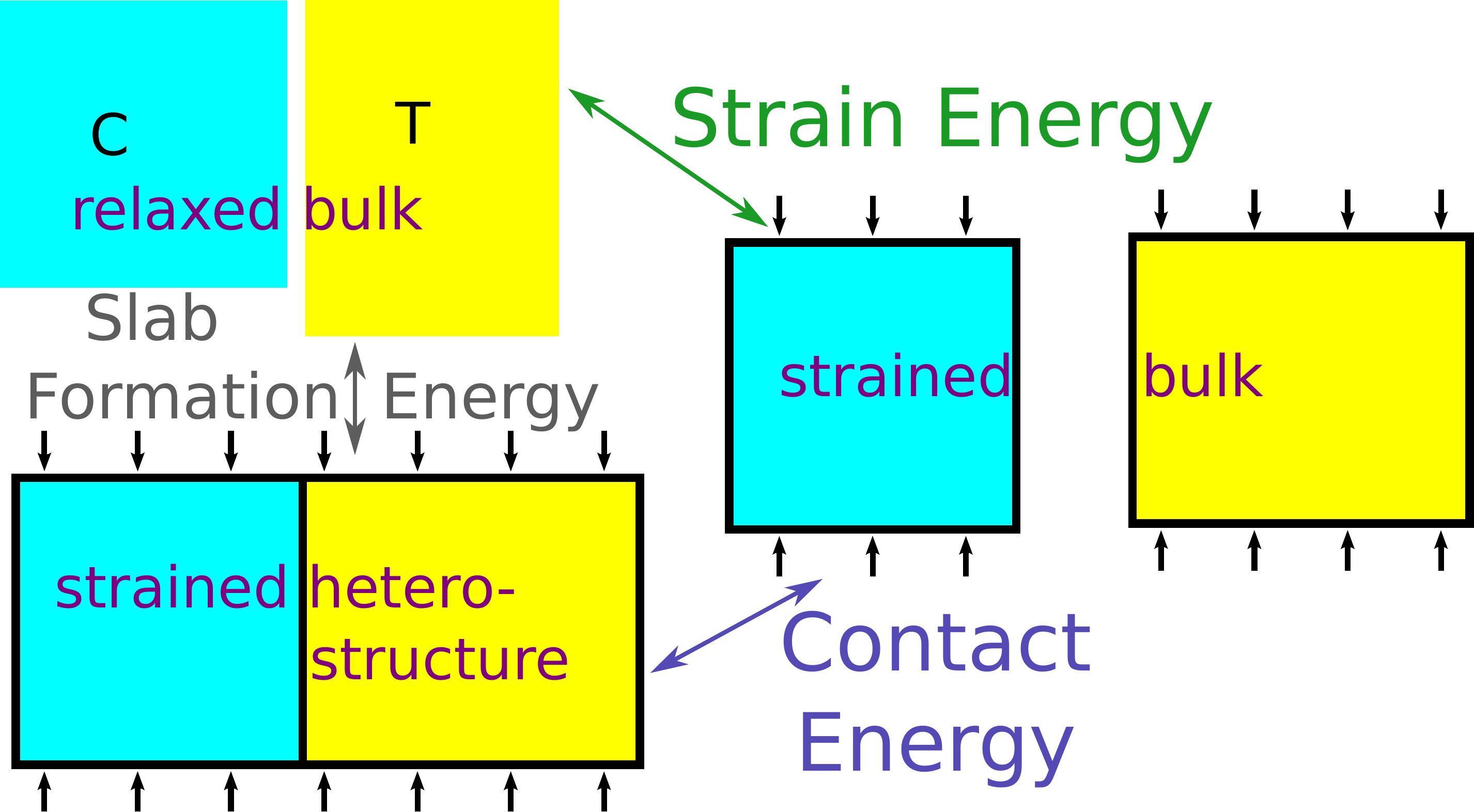} };
    \node[anchor=south east] (a) at +(-0.7, 0.3) { (a) };
  \end{tikzpicture}
  \\
  \includegraphics[width=0.47\textwidth]{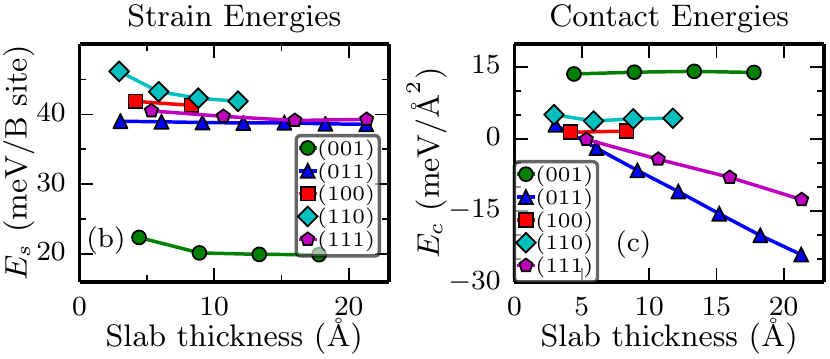}  
  \caption{
  (a) Three types of ZMGO structures used to decompose slab formation energies.
Black arrows represent cross-sectional biaxial strain, calculated by fully
 relaxing the strained heterostructure.
  All calculations use periodic boundary conditions.
  (b-c)
  Slab formation-energy decomposition, as defined in the text. 
  Motivation for unit choice is described in the text;
for comparison, note that a cubic unit cell has 16 B-sites and a
 (100)-cross-sectional area of approximately $74 \textrm{\AA}^2$.
Negative contact energy, as in the \{011\} surface, indicates energetic
preference for the strained layered-slab heterostructure over the strained
 bulk.
  }
  \label{fig:decomposition_energies}
\end{figure}
}

These energies are shown in Figure \ref{fig:decomposition_energies}(b-c). 
We note that \{111\} and especially \{011\} have negative contact energies,
meaning the strained layered-slab heterostructure is more stable than the
 strained ZGO and ZMO separated bulk.
Hence the local ionic distortions couple beneficially to the lattice strain,
i.e. atomic rearrangements partially alleviate the energy penalties of lattice
 strain.
This is predominantly concentrated in the breathing and tetragonal-distortion
local modes ($q_1$ and $q_3$ in the notation of Ref.
 \onlinecite{Vleck1939JCP}). %
Remarkably, the contact energy does not converge in the accessible thicknesses,
 due to the long range of the Jahn-Teller effect.
Rather, even a remarkable distance from the surface, interoctahedron coupling
 leads to intracellular atomic displacement.
Therefore the \{011\} preference originates in a beneficial coupling between
 strain and long-range atomic displacements
(contrary to previous non-first-principles work neglecting the latter
 \cite{Bouar1998AM,Ni2007AM,Ni2008AM,Ni2009NL,Ni2009NM}).

}

In conclusion, we have presented the physics of nanocheckerboards based on
 first-principles calculations.
We established that the thermodynamic ground state is complete phase separation.
The incomplete separation originates in diffusion limitations, leading to
 nanoscale phase domains.
We explained the observed cubic crystal structure at $x_{\mathrm{Mn}} = 0.25$
 based on noncooperative Jahn-Teller distortions at room temperature.
Therefore, although ZMGO's ground state is bulk-incoherent, the
diffusion-limited structure is bulk-coherent (using the terminology of Ref.
 \onlinecite{Liu2008PRB}).
This bulk coherence leads to phase separation of cubic and tetragonal phases
 along \{011\} surfaces, which we showed from first principles.
This, in the presence of kinetic constraints, automatically leads to
 checkerboards, based on pure geometry.
The preference for \{011\} surfaces originates in beneficial coupling between
 local distortions and lattice strain.
Further quantitative understanding will require robust models for the
 Jahn-Teller effect in doped materials at finite temperatures.

\begin{acknowledgments}
This research used resources of the National Energy Research Scientific
Computing Center, a DOE Office of Science User Facility supported by the Office
of Science of the U.S. Department of Energy under Contract No.
 DE-AC02-05CH11231.
The authors acknowledge support from a DARPA Young Faculty Award, Grant No.
 D13AP00051.
\end{acknowledgments}

%%%%%%%%%% Merge with supplemental materials %%%%%%%%%%
\pagebreak
\widetext
\begin{center}
\textbf{\large Supplemental Materials}
\end{center}
%%%%%%%%%% Merge with supplemental materials %%%%%%%%%%
%%%%%%%%%% Prefix a "S" to all equations, figures, tables and reset the counter %%%%%%%%%%
\setcounter{equation}{0}
\setcounter{figure}{0}
\setcounter{table}{0}
\setcounter{page}{1}
\makeatletter
\renewcommand{\theequation}{S\arabic{equation}}
\renewcommand{\thefigure}{S\arabic{figure}}
%%%%%%%%%% Prefix a "S" to all equations, figures, tables and reset the counter %%%%%%%%%%

\noindent
Here we describe various calculation details in the Letter.
\begin{itemize}
\item We provide lattice constants for various crystals according to both
 experiment and theory.
\item We detail the cluster expansion (CE) and related phase-diagram
calculations. We discuss the quality of the CE and its challenges for
 quantitative prediction.
  \item We justify GGA's calculation of a high-spin Mn electronic state.
\item We provide the calculation details for our expansion of surface energies
 in the symmetry-adapted harmonics.
\item We explain more details of the decomposition of multilayer-slab formation
 energies into strain and contact components.
\end{itemize}

{

\section{Lattice Constants} 

Here we detail the lattice constants of the bulk ZMO, ZGO, and mixed structure
 ($\mathrm{T}'$).
Data are provided in Table \ref{tab:lattices}.
DFT accurately captures the lattice constants of ZGO, ZMO, and the 
high-temperature fully-mixed ZMGO ($\mathrm{T}'$).
We model $\mathrm{T}'$ as the average of all calculated structures.
These calculations agree with experiment
 \cite{Errandonea2009PRB,Asbrink1999PRB,Yeo2006APL}
, within the 1-2\% overestimation characteristic of GGA calculations.
We adopt the convention where a cubic crystal has $c/a=1$, rather than
 $\sqrt{2}$.

{

\begin{table}[tbh]
  \centering
  \begin{tabular}{r l | c c c}
  % Generated by hightemp.py
Lattice param. &  &  Exp.      & DFT      &  DFT Err. \\
\hline
ZGO bulk      & $a$         &  8.34\AA &  8.46\AA &   1.5\% \\
ZMO bulk      & $a$         &  8.09\AA &  8.15\AA &   0.8\% \\
ZMGO hi-T     & $a$         &   8.2\AA &  8.31\AA &   1.3\% \\
ZGO$\to$ZMGO & $\delta a$  &  -1.6\% &  -1.8\% &          \\
ZMO$\to$ZMGO & $\delta a$  &   1.4\% &   1.9\% &          \\
\hline
ZGO bulk ($a$=) & $c$         &  8.34\AA &  8.46\AA &   1.5\% \\
ZMO bulk      & $c$         &  9.24\AA &  9.41\AA &   1.8\% \\
ZMGO hi-T     & $c$         &   8.7\AA &  8.89\AA &   2.2\% \\
ZGO$\to$ZMGO & $\delta c$  &   4.4\% &   5.1\% &            \\
ZMO$\to$ZMGO & $\delta c$  &  -5.9\% &  -5.5\% &            \\
\hline
ZMO bulk      & $c/a$       &  1.14    &  1.15    &   0.9\% \\
ZMGO hi-T     & $c/a$       &   1.06    &  1.07    &   0.9\% \\ 

  \end{tabular}
\caption{Lattice constants. Experimental data from Ref.
 \onlinecite{Errandonea2009PRB,Asbrink1999PRB,Yeo2006APL}.
}
  \label{tab:lattices}
\end{table} 
}

Table \ref{tab:rotations} compares theory and experiment for the rotation angle
 $\theta$ of the cubic nanochecker domains.
(See the Letter for the equation used.) 
It is apparent that this simple model successfully calculates the rotation to
 within $<1^\circ$.

{
\begin{table}
  \centering
  \begin{tabular}{c | c || c | c | c || c}
Ref. & Substrate & $\theta$ meas. &  $a$ (\AA) & $c$ (\AA) & $\theta_c =
 90^\circ - 2 \tan^{-1} a/c$ (calc.) \\
    \hline
\onlinecite{Yeo2006APL} & None & $\sim 6^\circ$ & $\approx 8.0$ & $\approx 9.0$
  & $6.7^\circ$ \\
\onlinecite{Park2008NL} & MgO & $5.2^\circ$ & $8.11$ & $8.95$ & $5.63^\circ$
 \\
\onlinecite{OMalley2008PRB} & MgO & $5.1^\circ$ & $8.14$ & $8.98$ &
 $5.62^\circ$
  \end{tabular}
\caption{Comparison of cubic-phase rotations $\theta$ between experiment and
 prediction. The simple equation is accurate to within $<1^\circ$.}
  \label{tab:rotations}
\end{table}
}

}{

\section{Cluster Expansion Calculations}

Here we detail the cluster expansion (CE) and related calculations used to
 calculate the phase diagram in Figure \ref{fig:phases}(b).
The CE was based on ZMGO structures. These calculations relied on VASP and the
 ATAT package; references appear in the Letter.

The supercells were generated with ATAT's MAPS code; then VASP fully relaxed
 the structure and calculated total energy.
Varying supercell sizes require varying numbers of k-points; we use a
$\Gamma$-centered mesh of at least 1000 k-points per reciprocal atom, as
 implemented in ATAT.
ATAT's MAPS code then performs a cluster expansion to these energies.
Then ATAT's EMC2 code uses the same cluster expansion for phase diagram
 calculations.

The structure generation, cluster determination, and effective cluster
interaction (ECI) fitting was performed with the default settings of ATAT's
 MAPS code.
The calculation used a total of 192 $\mathrm{ZnMn_xGa_{2-x}O_4}$ supercells, of
 size $\le 42$ atoms ($\approx 460 \textrm{\AA}^3$).
The expansion's cross-validation score is $1.1$ meV/B-site (compare to energies
 of formation of 0-80 meV/B).
A few additional structures relaxed to a high-energy peculiarity:
5 antiferromagnetic, %
4 low-spin, %
2 with reoriented distortions (some or all octahedra distorted in $y$ instead
 of $z$), %
and 2 anomalous cubic high-spin structures.  %
All remained above the convex hull.
When we restricted the antiferromagnetic and low-spin structures to a
 ferromagnetic high-spin configuration, the energy cost was only 4-12 meV/B.
These are omitted from the expansion, because they simply add noise to the
 physics being fit in the expansion.
Due to their high energy and minority, they cannot be expected to change the
 physics of the system.

Figure \ref{fig:eci}(a) shows the quality of the fit for the various computed
 structures.
Figure \ref{fig:eci}(b) shows the ECI as a function of cluster diameter. 
Intriguingly, the strongest interaction corresponds to the $V_{02}$ of
Wojtowicz \cite{Wojtowicz1959PR}, a pair repulsion between two adjacent atoms
 whose shared plane contains the JT-distorted direction.
We suspect this interaction's strength originates in the effect one cation's
$q_3$ (JT-active) mode has on its neighbors' octahedral modes and the
 associated energy cost.
However, such a discussion is beyond the scope of this work. 

{
\begin{figure}[tbh] 
  \includegraphics[width=0.80\textwidth]{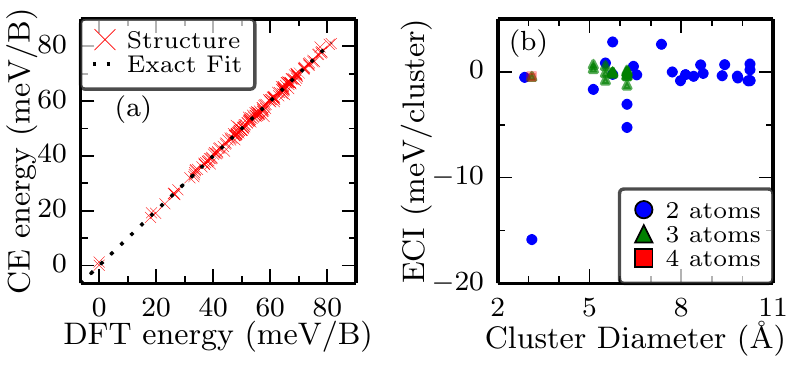}
  \caption{
(a) Quality of fit in ATAT-generated structures. The CE energies and DFT
 energies match well.
(b) Plot of ECI magnitudes as a function of distance. The picked CE used pairs,
 triplets, and quadruplets.
  }
  \label{fig:eci}
\end{figure}
}

We then computed the phase diagram using ATAT's EMC2 code.
We sampled temperatures from 100 to 1400 K in steps of 100 K, with chemical
 potential (differences) $\mu$ from -30 to +30 meV/B in steps of 0.5 meV/B.
For temperatures above 600 K, we additionally sampled $\mu$ from -120 to +120
 meV/B in steps of 0.5 meV/B and from -200 to +200 meV/B in steps of 5 meV/B.
Results appear in Figure \ref{fig:phases}(b) of the Letter.
We verified the phase diagram with DFT+U, where $U_\mathrm{Mn}=4$ eV, using the
 same VASP + MAPS + EMC2 calculations.
The phase diagram emerged qualitatively similar, although obviously the
 temperature scale differed.

As mentioned in the Letter, long-range effects are notoriously difficult to
 capture with the CE.
Later in the Letter, we analyze ZGO/ZMO slabs of thickness $t$ layered in the
 direction $\vec{\mathbf{k}}$.
Figure \ref{fig:faces_ce} compares the energies of these slabs according to DFT
 and the above CE.

{
\begin{figure}[tbh] 
  \includegraphics[width=0.67\textwidth]{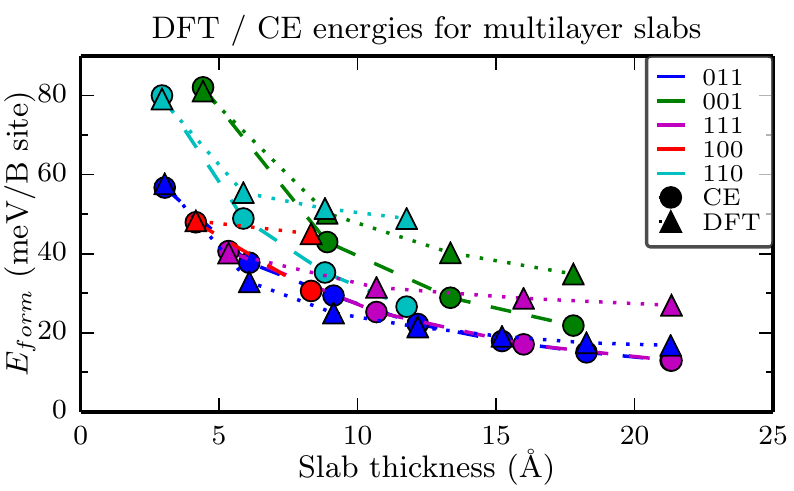}
  \caption{ 
Comparison of energies of multilayer ZGO/ZMO slabs, oriented in a particular
 direction, according to DFT and the cluster expansion (CE).
  }
  \label{fig:faces_ce}
\end{figure}
}

From this figure, we conclude:

\begin{enumerate}
  \item 
    The CE has a tendency to overmix relative to the exact result.
This is because the phase-separation is due to the long-range JT effect; once
the CE has difficulty capturing the full JT effect (see below), it will
 underestimate phase separation.
Therefore, given that the CE phase diagram shows total phase separation,
 \emph{a fortiori} this is the correct thermodynamic conclusion.
  \item 
The CE's failure to precisely resolve these energies means the expansion is
 insufficiently robust to perform first-principles simulations of kinetics.
    A more accurate model will be necessary for that.
\end{enumerate}

The origin of the CE's failure, as mentioned, lies in the long-range nature of
 the JT effect.
This is corroborated by the long-range contact energies described in the Letter.
Although the CE is a complete basis, in practice it is truncated after a finite
 number of terms.
To successfully capture this, we would need to include many long-range and
 large (i.e. many-atom) clusters to the expansion.

We attempted to capture these energetics with a variety of methods.

First, we added these multilayer slabs to the CE's list of known structures.
 However, the expansion still did not converge well.

Second, we attempted including reciprocal-space clusters in our fit, via the
 mixed-basis cluster expansion \cite{Laks1992PRB,Holliger2011PRB}.
This method fits energies to clusters in reciprocal space that are defined by
 the structure factor:
\begin{equation}
  S_{\vec{k}} = \sum_i \sigma_i \exp(\vec{r}_i\cdot\vec{k})
  \label{eq:mbce}
\end{equation}
where $\sigma_i = \pm 1$ refers to the pseudospin of a particular site and
 $\vec{r}_i$ its location.
However, this approach did not succeed either.
The variety of necessary $k$-space clusters, especially for the low-symmetry
 $I4_1/amd$ crystal, makes this difficult.
Additionally, the discrete phase boundaries make the $k$-space expansion
 difficult.

Third, we tried enhancing the fitting procedure with a compressive-sensing
approach similar to that of Ref.
 \onlinecite{Nelson2013PRB035125,Nelson2013PRB155105}.
This approach adds a penalty term for nonzero ECI in order to truncate the
 expansion after a few significant clusters.
However, practically this penalty term takes the form of the $\ell_1$ norm
(proportional to the ECI magnitude), rather than the $\ell_0$ norm (equal for
 all nonzero ECI).
This leads to homogenization of ECI magnitudes, rather than physically expected
 decay of ECI with cluster size and distance.
Despite attempts to compensate for this (similar to the reweighted
normalization in \cite{Nelson2013PRB155105}), we could not find a convergent
 cluster expansion for this difficult system.

Therefore, our CE is insufficiently robust for quantitative simulation, as
evidenced by its failure to successfully predict the energies of ZGO/ZMO
 multilayer slabs.
However, despite its tendency to overmix, it demonstrates the tendency of ZMGO
 to phase-separate even at high temperatures.

}{

\section{Spin-Crossover Calculations}
It may be tempting to propose a spin crossover transition, either to explain
the cubic structure at $x_{\mathrm{Mn}}\le 0.25$, or to cast aspersions on the
 GGA calculations.
Perhaps Ga-doping raises the crystal field splitting (by shrinking crystal
volume), moving the Mn's $d$-band occupation from $\ket{t_{2g\uparrow }^3
 e_{g\uparrow}^1}$ to $\ket{t_{2g\uparrow}^3t_{2g\downarrow}^1}$.
Then the insignificantly weak JT effect in the $\ket{t_{2g}}$ manifold leads to
 an undistorted crystal strucutre.
In fact, this spin-crossover transition was attributed to the
tetragonal-to-cubic transition of ZMO at a pressure of 23 GPa
 \cite{Asbrink1999PRB,Choi2006PRB,Yamanaka2008AMin}.
Therefore, we further scrutinize GGA's prediction of spin state and its
 reliability.

{
\begin{figure}[tbh] 
  \includegraphics[width=0.57\textwidth]{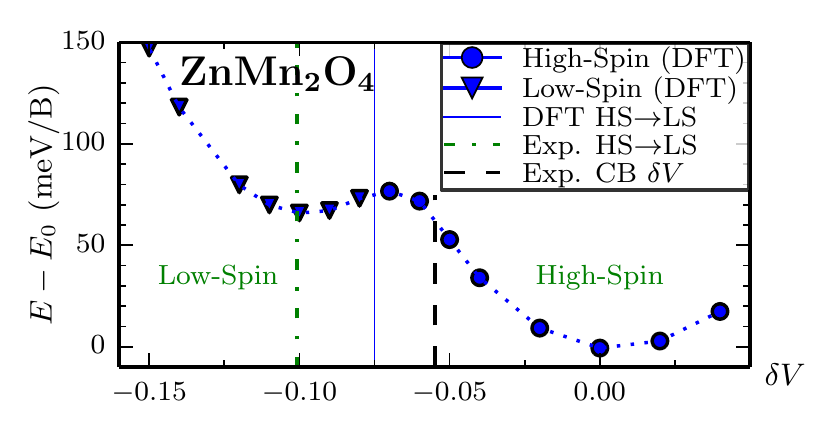}
  \caption{ 
  ZMO's spin-crossover transition.
Energies (relative to the high-spin fully-relaxed ground state) are plotted as
 a function of fractional volume change.
We show the checkerboard (CB) structure is in the high-spin regime according to
 GGA and \emph{a fortiori} in experiment.
  }
  \label{fig:sco}
\end{figure}
}

However, it is clear that GGA is correct in predicting a high-spin state.
As illustrated in Figure \ref{fig:sco}, the reported nanocheckerboard lattice
parameters place only a 5\% (volumetric) strain on the Mn octahedra, whereas
the ZMO phase transition occurs at a 10\% (volumetric) strain
 \cite{Asbrink1999PRB};
furthermore, GGA predicts the ZMO transition at only 7\% (volumetric) strain.
It is obvious, then, that GGA has a tendency to overpredict the low-spin state,
 relative to experiment.
Therefore, whereas GGA overpredicts the spin-crossover transition in ZMO, the
absence of such a prediction in our structures---and the insufficient strain
for such a transition---indicates that there is no spin-crossover transition in
 our structures or checkerboards.
According to this, experiment should find high-spin Mn in $\mathrm{Zn Mn_{0.5}
 Ga_{1.5} O_4}$, in agreement with GGA;
no such experiments have been reported to our knowledge.

}{

\section{Symmetry-Adapted Harmonics}

In the Letter, we present formation energies for multilayer ZGO/ZMO slabs
 oriented in five directions.
Here we detail the expansion to slabs oriented in an arbitrary direction.

It is obvious from Figure \ref{fig:slab_energies}(b) that the slab formation
energy of multilayer slabs of thickness $t$ and cross-section $A$, oriented in
 direction $\vec{\mathbf{k}}$, can be expressed as
$E(\vec{\mathbf{k}},t) = A( c_a( \vec{\mathbf{k}}) + t c_v (\vec{\mathbf{k}}))$
, where $c_a$ and $c_v$ scale with area and volume ($= A t$), respectively.
The values of $c_a,c_v$ for five directions are calculated from first
 principles as shown in Figure \ref{fig:slab_energies}(b).

We now expand the coefficients $c(\vec{\mathbf{k}})$ as linear combinations of
 the symmetry-adapted harmonics, here the tetragonal ($D_{4h}$) harmonics.
These polynomials are orthonormal, where we take the inner product as:
\begin{equation}
\langle f(\vec{\mathbf{k}}) \vert g(\vec{\mathbf{k}}) \rangle = \int_{\vert
\vec{\mathbf{k}}\vert = 1}  \frac{f(\vec{\mathbf{k}}) g(\vec{\mathbf{k}}) }{
\pi^2} dS = \int_0^\pi d\theta \int_0^{2\pi} d\varphi \frac{f(\theta,\varphi)
 g(\theta,\varphi) }{ \pi^2}
\end{equation}

{
\begin{table}[h]
  \centering
  \begin{tabular}{c | c | c | c | c}
$f_n(\theta,\varphi)$ & $\frac{1}{\sqrt{2}} $ & $\cos 2\theta$ & $\cos 4\theta$
 & $\sqrt{\frac{128}{35}} \cos 4\varphi \sin^4 \theta$  \\
    &
    \includegraphics[height=36pt]{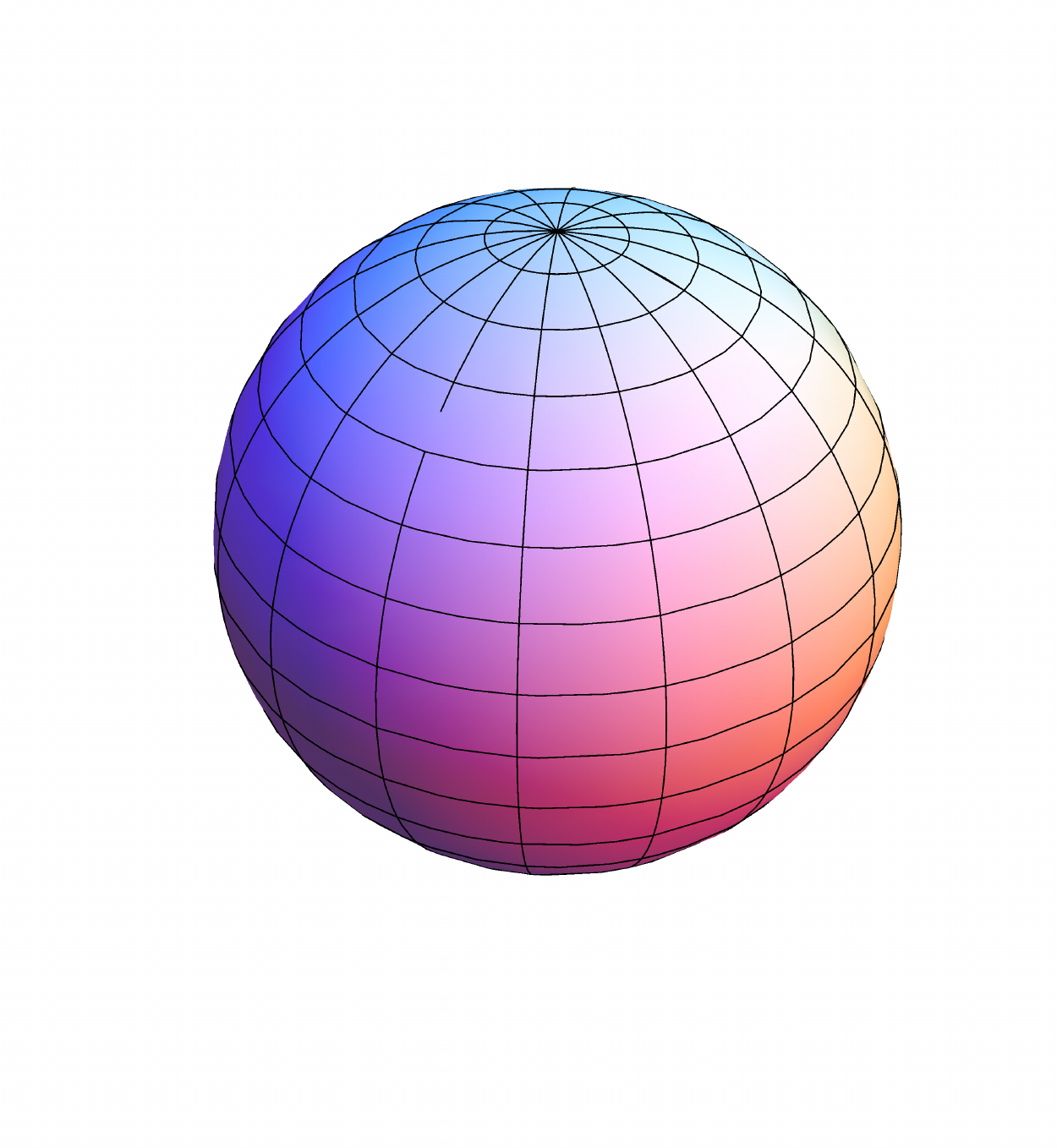}  &
    \includegraphics[height=36pt]{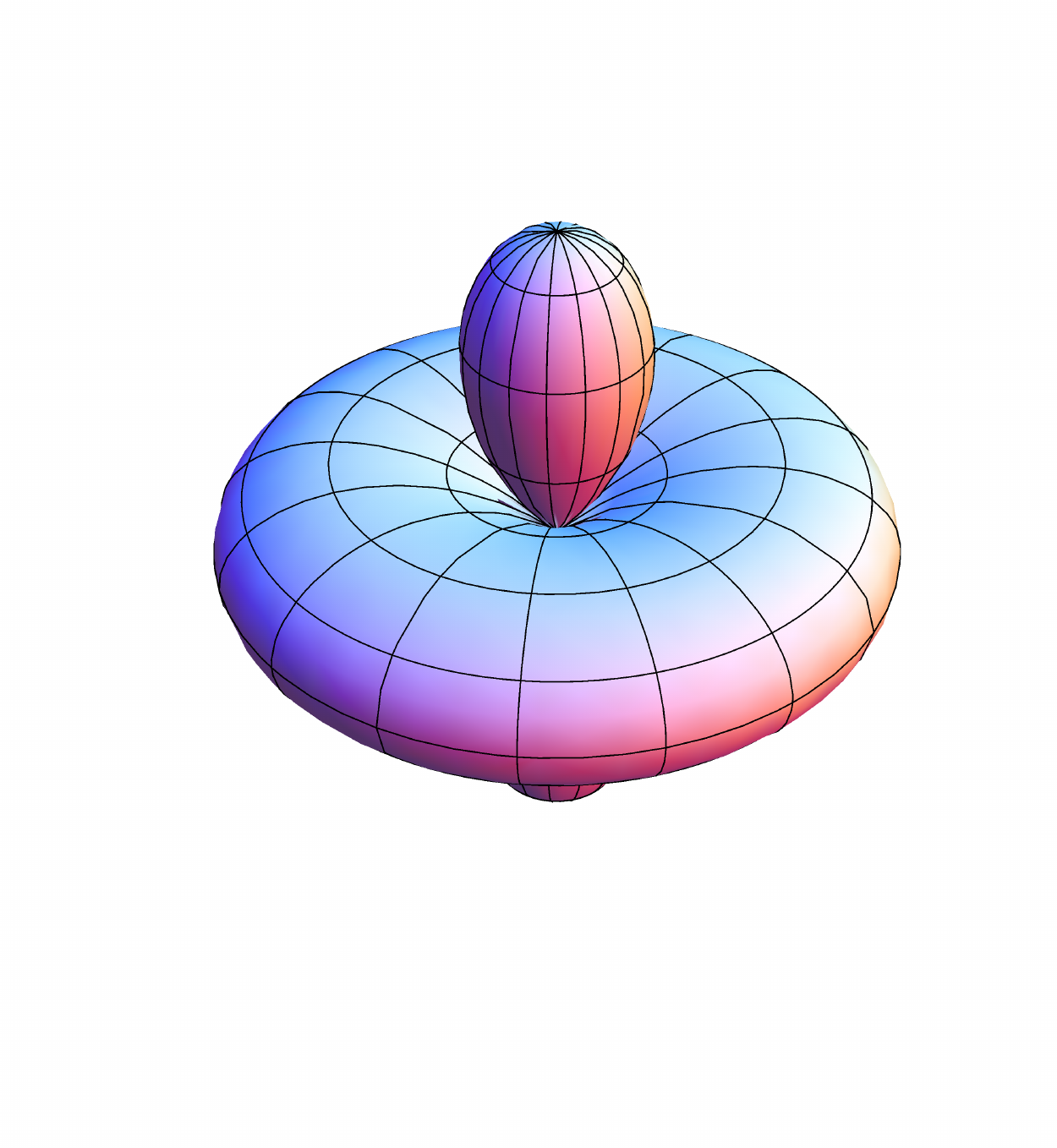}  &
    \includegraphics[height=36pt]{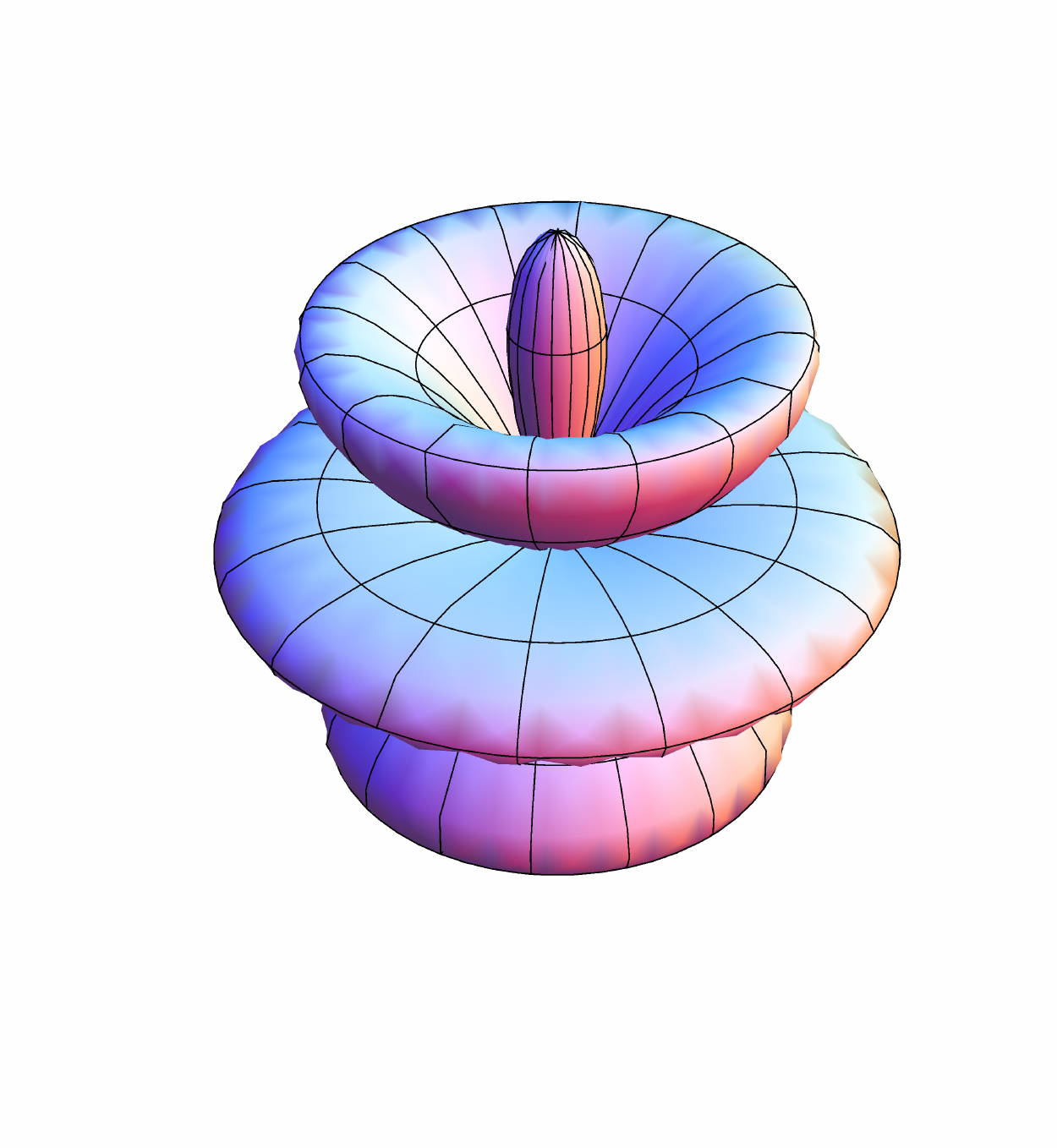}  &
    \includegraphics[height=36pt]{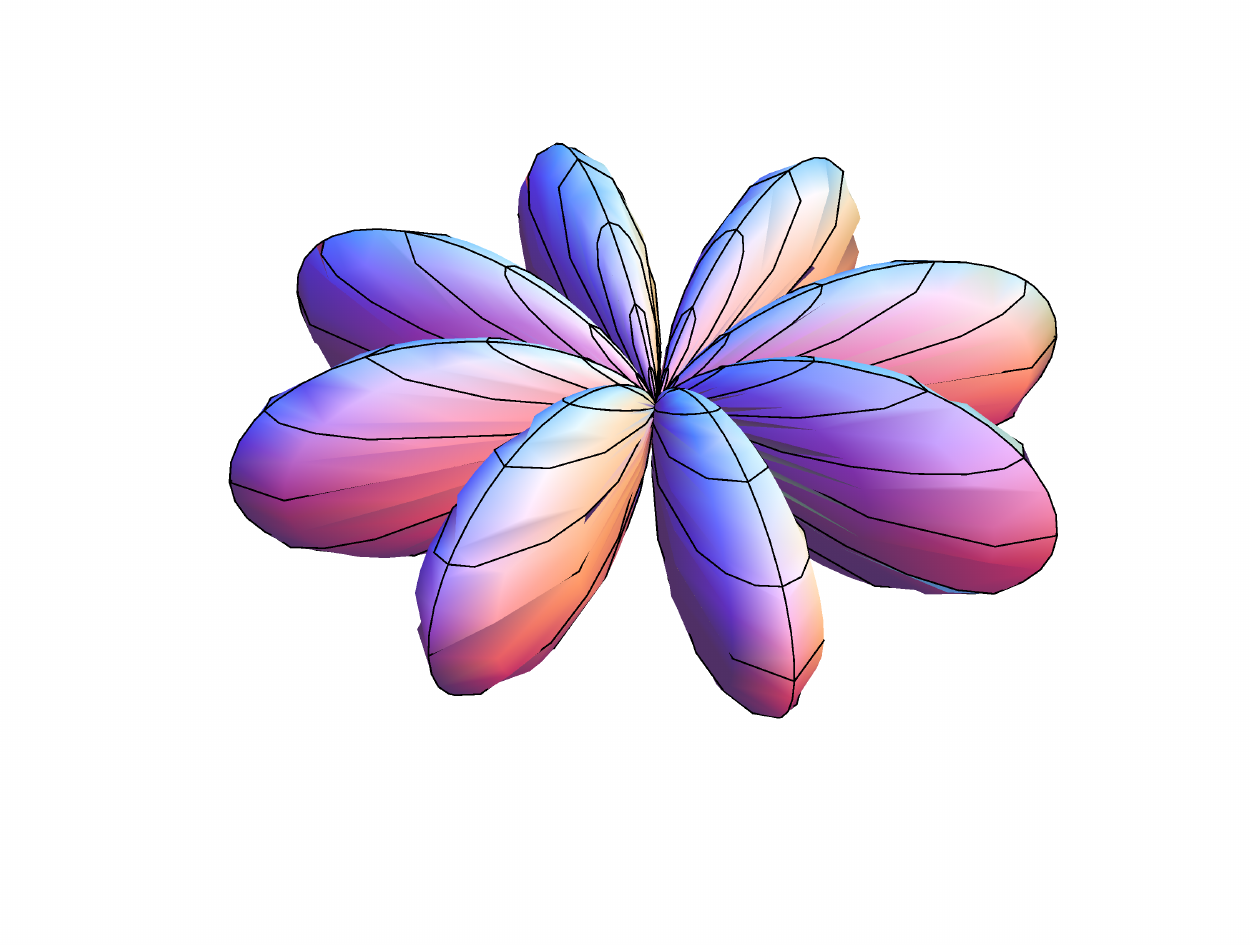}  \\
    \hline
    Order & $x^0$ & $x^2$ & $x^4$ & $x^4$ \\
    Irrep. & $A_{1g}$ & $A_{1g}$ & $A_{1g}^2$ & $B_{1g}^2$  \\
$c_a$ (eV/$\textrm{\AA}^2$) & $1\times 10^{-2}$ & $6\times 10^{-3}$ & $1 \times
 10^{-3}$ & $-9 \times 10^{-4}$ \\
$c_v$ (eV/$\textrm{\AA}^3$)& $2\times 10^{-3}$ & $-6 \times 10^{-4}$ & $5
 \times 10^{-4}$ & $ 8 \times 10^{-6}$
  \end{tabular}
  \caption{
  Expansion of slab formation energies in symmetry-adapted harmonics.
  }
  \label{tab:harmonics}
\end{table}
}

Table \ref{tab:harmonics} shows the first four terms, along with a graphical
representation, term order, and irreducible representation that the term
 transforms as.
(The polynomial transforming as $B_{2g}^2$ also is of the order of $x^4$, but
 is a linear combination of those listed.)
The same table lists values for $c_a$ and $c_v$, obtained by a least-squares
 fit of the four expansion terms to the five known energies.

}{
\iftrue %

\section{Energy Decomposition}  %

In the Letter, we describe the energy decomposition into strain and contact
 energy, and specifically its interpretation of the (011) preference.
Here we present this decomposition in more detail.

The formation energy of a $\mathrm{ZnMn_xGa_{2-x}O_4}$ structure (where
 $x_{\mathrm{Mn}} = x/2$) is:
\begin{equation}
\varepsilon_{\mathrm{form}} = \varepsilon_{\mathrm{tot}} - x_{\mathrm{Mn}}
 \varepsilon_{\mathrm{ZMO}} - (1 - x_{\mathrm{Mn}}) \varepsilon_{\mathrm{ZGO}}
  \label{eq:enform}
\end{equation}
where all energies $\varepsilon$ are given per-B-site, and
 $\varepsilon_\mathrm{ZG(M)O}$ refers to the energy of bulk ZG(M)O.
The energy of formation is thus the difference between the energy and the
 ``tie-line'' connecting the ZGO / ZMO extrema.

The energy of multilayer slabs are comprised of
(a) bulk energies of the ZGO and ZMO slabs;
(b) strain energy due to coherent lattice matching;
(c) chemical binding energies; 
and (d) intracellular atomic displacements near the surface
 ($\vec{\mathbf{k}}=0$ optical phonons).
We combine these last two into a contact energy, noting that (unlike the
 strain) it must decay for infinitely thick slabs.
We avoid the term ``surface energy'' because it can ambiguously refer to
 contact energy or total slab formation energy.

Therefore, for two slabs of thickness $t$ joined in direction
 $\hat{\mathbf{k}}$:
\begin{equation}
E_{tot}=N (x_{\mathrm{Mn}} \varepsilon_{\mathrm{ZMO}} + (1-x_{\mathrm{Mn}})
\varepsilon_{\mathrm{ZGO}}) + N E_{strain}(\hat{\mathbf{k}},t) + A
 E_{contact}(\hat{\mathbf{k}},t)
\end{equation}

We now seek to determine each of these terms from first-energy calculations. 
Consider the three structures shown in Figure \ref{fig:decomposition}:
relaxed bulk (RB) of pure ZGO and ZMO, strained heterostructure (SH) of
fully-relaxed adjacent slabs, and strained bulk (SB) of pure ZM(G)O with
 lattice parameters set to those of SH.

{
\begin{figure}[tbh]
  \includegraphics[width=0.50\textwidth]{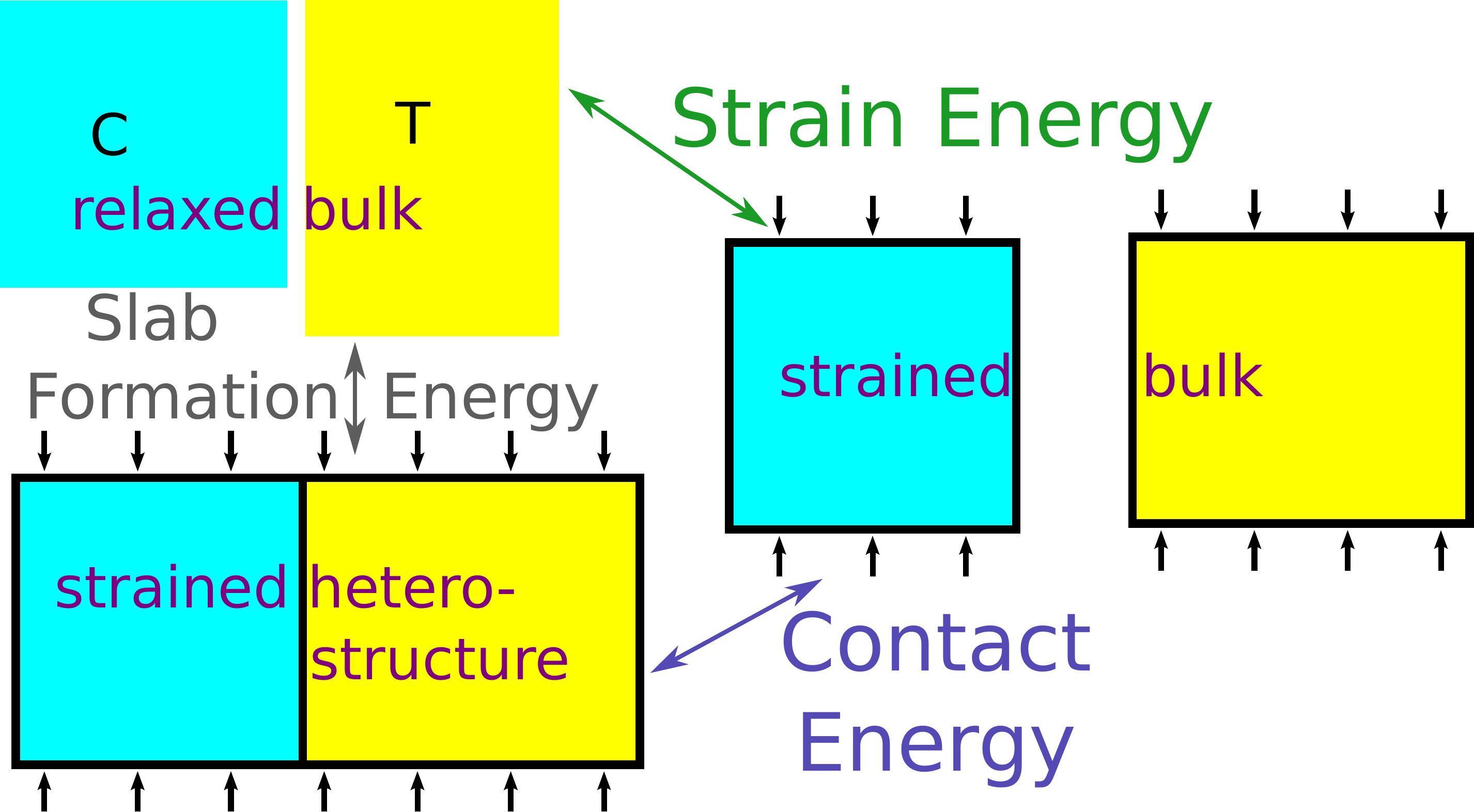} \\
  \caption{
  Three types of ZMGO structures used to decompose slab formation energies.
Arrows represent cross-sectional biaxial strain, calculated by fully relaxing
 the heterostructure.
  All calculations use periodic boundary conditions.
  See text for discussion.
  }
  \label{fig:decomposition}
\end{figure}
}

We can easily calculate the energy of each structure with DFT, using periodic
 boundary conditions.
It is immediately apparent that
\begin{align}
E_{RB} &= N (x_{\mathrm{Mn}} E_{\mathrm{ZMO}} + (1-x_{\mathrm{Mn}})
 E_{\mathrm{ZGO}}) \\
E_{SB} &= N (x_{\mathrm{Mn}} E_{\mathrm{ZMO}} + (1-x_{\mathrm{Mn}})
 E_{\mathrm{ZGO}}) + N E_{strain}(\hat{\mathbf{k}},t)  \\
E_{SH} &= N (x_{\mathrm{Mn}} E_{\mathrm{ZMO}} + (1-x_{\mathrm{Mn}})
E_{\mathrm{ZGO}}) + N E_{strain}(\hat{\mathbf{k}},t) + A
 E_{contact}(\hat{\mathbf{k}},t)
\end{align}
It is then trivial to extract $E_{strain}$ and $E_{contact}$ from DFT energies,
 as depicted in Figure \ref{fig:decomposition}.

Results are presented in the Letter (Fig. \ref{fig:decomposition_energies}).
For example, (001) layering 
contains little strain energy, because the JT-distorted [001] direction is
perpendicular to the surface, so ZGO can achieve a cubic lattice and ZMO a
 tetragonal one.
However, this is offset by significant contact energy, due to atomic
 rearrangements (interoctahedron coupling) near the surface.

In the Letter we interpret the (011) preference with this decomposition.

\fi 
}

\bibliography{current}

%Merlin.mbs v4.21 2009-07-09.
\begin{thebibliography}{10}%
\makeatletter
\providecommand \@ifxundefined [1]{%
 \ifx #1\undefined \expandafter \@firstoftwo
 \else \expandafter \@secondoftwo
\fi
}%
\providecommand \@ifnum [1]{%
 \ifnum #1\expandafter \@firstoftwo
 \else \expandafter \@secondoftwo
\fi
}%
\providecommand \enquote [1]{``#1''}%
\providecommand \bibnamefont  [1]{#1}%
\providecommand \bibfnamefont [1]{#1}%
\providecommand \citenamefont [1]{#1}%
\providecommand\href[0]{\@sanitize\@href}%
\providecommand\@href[1]{\endgroup\@@startlink{#1}\endgroup\@@href}%
\providecommand\@@href[1]{#1\@@endlink}%
\providecommand \@sanitize [0]{\begingroup\catcode`\&12\catcode`\#12\relax}%
\@ifxundefined \pdfoutput {\@firstoftwo}{%
 \@ifnum{\z@=\pdfoutput}{\@firstoftwo}{\@secondoftwo}%
}{%
 \providecommand\@@startlink[1]{\leavevmode\special{html:<a href="#1">}}%
 \providecommand\@@endlink[0]{\special{html:</a>}}%
}{%
 \providecommand\@@startlink[1]{%
  \leavevmode
  \pdfstartlink
   attr{/Border[0 0 1 ]/H/I/C[0 1 1]}%
   user{/Subtype/Link/A<</Type/Action/S/URI/URI(#1)>>}%
  \relax
 }%
 \providecommand\@@endlink[0]{\pdfendlink}%
}%
\providecommand \url  [0]{\begingroup\@sanitize \@url }%
\providecommand \@url [1]{\endgroup\@href {#1}{\urlprefix}}%
\providecommand \urlprefix [0]{URL }%
\providecommand \Eprint[0]{\href }%
\@ifxundefined \urlstyle {%
  \providecommand \doi [1]{doi:\discretionary{}{}{}#1}%
}{%
  \providecommand \doi [0]{doi:\discretionary{}{}{}\begingroup
  \urlstyle{rm}\Url }%
}%
\providecommand \doibase [0]{http://dx.doi.org/}%
\providecommand \Doi[1]{\href{\doibase#1}}%
\providecommand \bibAnnote [3]{%
  \BibitemShut{#1}%
  \begin{quotation}\noindent
    \textsc{Key:}\ #2\\\textsc{Annotation:}\ #3%
  \end{quotation}%
}%
\providecommand \bibAnnoteFile [2]{%
  \IfFileExists{#2}{\bibAnnote {#1} {#2} {\input{#2}}}{}%
}%
\providecommand \typeout [0]{\immediate \write \m@ne }%
\providecommand \selectlanguage [0]{\@gobble}%
\providecommand \bibinfo [0]{\@secondoftwo}%
\providecommand \bibfield [0]{\@secondoftwo}%
\providecommand \translation [1]{[#1]}%
\providecommand \BibitemOpen[0]{}%
\providecommand \bibitemStop [0]{}%
\providecommand \bibitemNoStop [0]{.\EOS\space}%
\providecommand \EOS [0]{\spacefactor3000\relax}%
\providecommand \BibitemShut [1]{\csname bibitem#1\endcsname}%
%</preamble>
\bibitem{Leroux1991PMB}%
  \BibitemOpen
  \bibfield{author}{%
  \bibinfo {author} {\bibfnamefont{C.}~\bibnamefont{Leroux}}, \bibinfo {author}
  {\bibfnamefont{A.}~\bibnamefont{Loiseau}}, \bibinfo {author}
  {\bibfnamefont{D.}~\bibnamefont{Broddin}},\ and\ \bibinfo {author}
  {\bibfnamefont{G.}~\bibnamefont{Vantendeloo}},\ }%
  \bibfield{journal}{%
  \Doi{10.1080/13642819108207603}{\bibinfo {journal} {Phil. Mag. B}}\ }%
  \textbf{\bibinfo {volume} {64}},\ \bibinfo {pages} {57} (\bibinfo {month}
  {Jul}\ \bibinfo {year} {1991})%
  \bibAnnoteFile{NoStop}{Leroux1991PMB}%
\bibitem{Udoh1995MSEA}%
  \BibitemOpen
  \bibfield{author}{%
  \bibinfo {author} {\bibfnamefont{K.-I.}\ \bibnamefont{Udoh}}, \bibinfo
  {author} {\bibfnamefont{A.~E.}\ \bibnamefont{Araby}}, \bibinfo {author}
  {\bibfnamefont{Y.}~\bibnamefont{Tanaka}}, \bibinfo {author}
  {\bibfnamefont{K.}~\bibnamefont{Hisatsune}}, \bibinfo {author}
  {\bibfnamefont{K.}~\bibnamefont{Yasuda}}, \bibinfo {author}
  {\bibfnamefont{G.~V.}\ \bibnamefont{Tendeloo}},\ and\ \bibinfo {author}
  {\bibfnamefont{J.~V.}\ \bibnamefont{Landuyt}},\ }%
  \bibfield{journal}{%
  \Doi{10.1016/0921-5093(95)09850-x}{\bibinfo {journal} {Mater. Sci. Eng. A}}\
  }%
  \textbf{\bibinfo {volume} {203}},\ \bibinfo {pages} {154} (\bibinfo {month}
  {Nov}\ \bibinfo {year} {1995})%
  \bibAnnoteFile{NoStop}{Udoh1995MSEA}%
\bibitem{Winn2000JAC}%
  \BibitemOpen
  \bibfield{author}{%
  \bibinfo {author} {\bibfnamefont{H.}~\bibnamefont{Winn}}, \bibinfo {author}
  {\bibfnamefont{Y.}~\bibnamefont{Tanaka}}, \bibinfo {author}
  {\bibfnamefont{T.}~\bibnamefont{Shiraishi}}, \bibinfo {author}
  {\bibfnamefont{K.}~\bibnamefont{Udoh}}, \bibinfo {author}
  {\bibfnamefont{E.}~\bibnamefont{Miura}}, \bibinfo {author}
  {\bibfnamefont{R.}~\bibnamefont{Hernandez}}, \bibinfo {author}
  {\bibfnamefont{Y.}~\bibnamefont{Takuma}},\ and\ \bibinfo {author}
  {\bibfnamefont{K.}~\bibnamefont{Hisatsune}},\ }%
  \bibfield{journal}{%
  \Doi{10.1016/s0925-8388(00)00791-x}{\bibinfo {journal} {J. Alloys Compd.}}\
  }%
  \textbf{\bibinfo {volume} {306}},\ \bibinfo {pages} {262} (\bibinfo {month}
  {Jun}\ \bibinfo {year} {2000})%
  \bibAnnoteFile{NoStop}{Winn2000JAC}%
\bibitem{Zheng2004Science}%
  \BibitemOpen
  \bibfield{author}{%
  \bibinfo {author} {\bibfnamefont{H.}~\bibnamefont{Zheng}},\ }%
  \bibfield{journal}{%
  \Doi{10.1126/science.1094207}{\bibinfo {journal} {Science}}\ }%
  \textbf{\bibinfo {volume} {303}},\ \bibinfo {pages} {661} (\bibinfo {month}
  {Jan}\ \bibinfo {year} {2004})%
  \bibAnnoteFile{NoStop}{Zheng2004Science}%
\bibitem{Yeo2006APL}%
  \BibitemOpen
  \bibfield{author}{%
  \bibinfo {author} {\bibfnamefont{S.}~\bibnamefont{Yeo}}, \bibinfo {author}
  {\bibfnamefont{Y.}~\bibnamefont{Horibe}}, \bibinfo {author}
  {\bibfnamefont{S.}~\bibnamefont{Mori}}, \bibinfo {author}
  {\bibfnamefont{C.}~\bibnamefont{Tseng}}, \bibinfo {author}
  {\bibfnamefont{C.}~\bibnamefont{Chen}}, \bibinfo {author}
  {\bibfnamefont{A.}~\bibnamefont{Khachaturyan}}, \bibinfo {author}
  {\bibfnamefont{C.}~\bibnamefont{Zhang}},\ and\ \bibinfo {author}
  {\bibfnamefont{S.-W.}\ \bibnamefont{Cheong}},\ }%
  \bibfield{journal}{%
  \Doi{10.1063/1.2402115}{\bibinfo {journal} {Appl. Phys. Lett.}}\ }%
  \textbf{\bibinfo {volume} {89}},\ \bibinfo {pages} {233120} (\bibinfo {year}
  {2006})%
  \bibAnnoteFile{NoStop}{Yeo2006APL}%
\bibitem{Park2008NL}%
  \BibitemOpen
  \bibfield{author}{%
  \bibinfo {author} {\bibfnamefont{S.}~\bibnamefont{Park}}, \bibinfo {author}
  {\bibfnamefont{Y.}~\bibnamefont{Horibe}}, \bibinfo {author}
  {\bibfnamefont{T.}~\bibnamefont{Asada}}, \bibinfo {author}
  {\bibfnamefont{L.}~\bibnamefont{Wielunski}}, \bibinfo {author}
  {\bibfnamefont{N.}~\bibnamefont{Lee}}, \bibinfo {author}
  {\bibfnamefont{P.}~\bibnamefont{Bonanno}}, \bibinfo {author}
  {\bibfnamefont{S.}~\bibnamefont{O'Malley}}, \bibinfo {author}
  {\bibfnamefont{A.}~\bibnamefont{Sirenko}}, \bibinfo {author}
  {\bibfnamefont{A.}~\bibnamefont{Kazimirov}}, \bibinfo {author}
  {\bibfnamefont{M.}~\bibnamefont{Tanimura}}, \emph{et~al.},\ }%
  \bibfield{journal}{%
  \Doi{10.1021/nl072848s}{\bibinfo {journal} {Nano Lett.}}\ }%
  \textbf{\bibinfo {volume} {8}},\ \bibinfo {pages} {720} (\bibinfo {year}
  {2008})%
  \bibAnnoteFile{NoStop}{Park2008NL}%
\bibitem{OMalley2008PRB}%
  \BibitemOpen
  \bibfield{author}{%
  \bibinfo {author} {\bibfnamefont{S.}~\bibnamefont{O'Malley}}, \bibinfo
  {author} {\bibfnamefont{P.}~\bibnamefont{Bonanno}}, \bibinfo {author}
  {\bibfnamefont{K.}~\bibnamefont{Ahn}}, \bibinfo {author}
  {\bibfnamefont{A.}~\bibnamefont{Sirenko}}, \bibinfo {author}
  {\bibfnamefont{A.}~\bibnamefont{Kazimirov}}, \bibinfo {author}
  {\bibfnamefont{M.}~\bibnamefont{Tanimura}}, \bibinfo {author}
  {\bibfnamefont{T.}~\bibnamefont{Asada}}, \bibinfo {author}
  {\bibfnamefont{S.}~\bibnamefont{Park}}, \bibinfo {author}
  {\bibfnamefont{Y.}~\bibnamefont{Horibe}},\ and\ \bibinfo {author}
  {\bibfnamefont{S.}~\bibnamefont{Cheong}},\ }%
  \bibfield{journal}{%
  \Doi{10.1103/PhysRevB.78.165424}{\bibinfo {journal} {Phys. Rev. B}}\ }%
  \textbf{\bibinfo {volume} {78}},\ \bibinfo {pages} {165424} (\bibinfo {year}
  {2008})%
  \bibAnnoteFile{NoStop}{OMalley2008PRB}%
\bibitem{Zhang2007APL133123}%
  \BibitemOpen
  \bibfield{author}{%
  \bibinfo {author} {\bibfnamefont{C.}~\bibnamefont{Zhang}}, \bibinfo {author}
  {\bibfnamefont{S.}~\bibnamefont{Yeo}}, \bibinfo {author}
  {\bibfnamefont{Y.}~\bibnamefont{Horibe}}, \bibinfo {author}
  {\bibfnamefont{Y.}~\bibnamefont{Choi}}, \bibinfo {author}
  {\bibfnamefont{S.}~\bibnamefont{Guha}}, \bibinfo {author}
  {\bibfnamefont{M.}~\bibnamefont{Croft}}, \bibinfo {author}
  {\bibfnamefont{S.-W.}\ \bibnamefont{Cheong}},\ and\ \bibinfo {author}
  {\bibfnamefont{S.}~\bibnamefont{Mori}},\ }%
  \bibfield{journal}{%
  \Doi{10.1063/1.2717568}{\bibinfo {journal} {Appl. Phys. Lett.}}\ }%
  \textbf{\bibinfo {volume} {90}},\ \bibinfo {pages} {133123} (\bibinfo {year}
  {2007})%
  \bibAnnoteFile{NoStop}{Zhang2007APL133123}%
\bibitem{Zhang2007APL233110}%
  \BibitemOpen
  \bibfield{author}{%
  \bibinfo {author} {\bibfnamefont{C.}~\bibnamefont{Zhang}}, \bibinfo {author}
  {\bibfnamefont{C.}~\bibnamefont{Tseng}}, \bibinfo {author}
  {\bibfnamefont{C.}~\bibnamefont{Chen}}, \bibinfo {author}
  {\bibfnamefont{S.}~\bibnamefont{Yeo}}, \bibinfo {author}
  {\bibfnamefont{Y.}~\bibnamefont{Choi}},\ and\ \bibinfo {author}
  {\bibfnamefont{S.}~\bibnamefont{Cheong}},\ }%
  \bibfield{journal}{%
  \Doi{10.1063/1.2821838}{\bibinfo {journal} {Appl. Phys. Lett.}}\ }%
  \textbf{\bibinfo {volume} {91}},\ \bibinfo {pages} {233110} (\bibinfo {year}
  {2007})%
  \bibAnnoteFile{NoStop}{Zhang2007APL233110}%
\bibitem{Bouar1998AM}%
  \BibitemOpen
  \bibfield{author}{%
  \bibinfo {author} {\bibfnamefont{Y.}~\bibnamefont{Le~Bouar}}, \bibinfo
  {author} {\bibfnamefont{A.}~\bibnamefont{Loiseau}},\ and\ \bibinfo {author}
  {\bibfnamefont{A.}~\bibnamefont{Khachaturyan}},\ }%
  \bibfield{journal}{%
  \Doi{10.1016/S1359-6454(97)00455-2}{\bibinfo {journal} {Acta Mater.}}\ }%
  \textbf{\bibinfo {volume} {46}},\ \bibinfo {pages} {2777} (\bibinfo {year}
  {1998})%
  \bibAnnoteFile{NoStop}{Bouar1998AM}%
\bibitem{Ni2007AM}%
  \BibitemOpen
  \bibfield{author}{%
  \bibinfo {author} {\bibfnamefont{Y.}~\bibnamefont{Ni}}, \bibinfo {author}
  {\bibfnamefont{Y.}~\bibnamefont{Jin}},\ and\ \bibinfo {author}
  {\bibfnamefont{A.}~\bibnamefont{Khachaturyan}},\ }%
  \bibfield{journal}{%
  \Doi{10.1016/j.actamat.2007.05.016}{\bibinfo {journal} {Acta Mater.}}\ }%
  \textbf{\bibinfo {volume} {55}},\ \bibinfo {pages} {4903} (\bibinfo {year}
  {2007})%
  \bibAnnoteFile{NoStop}{Ni2007AM}%
\bibitem{Ni2008AM}%
  \BibitemOpen
  \bibfield{author}{%
  \bibinfo {author} {\bibfnamefont{Y.}~\bibnamefont{Ni}}\ and\ \bibinfo
  {author} {\bibfnamefont{A.~G.}\ \bibnamefont{Khachaturyan}},\ }%
  \bibfield{journal}{%
  \Doi{10.1016/j.actamat.2008.05.035}{\bibinfo {journal} {Acta Mater.}}\ }%
  \textbf{\bibinfo {volume} {56}},\ \bibinfo {pages} {4498} (\bibinfo {year}
  {2008})%
  \bibAnnoteFile{NoStop}{Ni2008AM}%
\bibitem{Ni2009NL}%
  \BibitemOpen
  \bibfield{author}{%
  \bibinfo {author} {\bibfnamefont{Y.}~\bibnamefont{Ni}}, \bibinfo {author}
  {\bibfnamefont{W.}~\bibnamefont{Rao}},\ and\ \bibinfo {author}
  {\bibfnamefont{A.~G.}\ \bibnamefont{Khachaturyan}},\ }%
  \bibfield{journal}{%
  \Doi{10.1021/nl901551j}{\bibinfo {journal} {Nano Lett.}}\ }%
  \textbf{\bibinfo {volume} {9}},\ \bibinfo {pages} {3275} (\bibinfo {month}
  {Sep}\ \bibinfo {year} {2009})%
  \bibAnnoteFile{NoStop}{Ni2009NL}%
\bibitem{Ni2009NM}%
  \BibitemOpen
  \bibfield{author}{%
  \bibinfo {author} {\bibfnamefont{Y.}~\bibnamefont{Ni}}\ and\ \bibinfo
  {author} {\bibfnamefont{A.~G.}\ \bibnamefont{Khachaturyan}},\ }%
  \bibfield{journal}{%
  \Doi{10.1038/nmat2431}{\bibinfo {journal} {Nat. Mater.}}\ }%
  \textbf{\bibinfo {volume} {8}},\ \bibinfo {pages} {410} (\bibinfo {month}
  {May}\ \bibinfo {year} {2009})%
  \bibAnnoteFile{NoStop}{Ni2009NM}%
\bibitem{Boettinger2002ARMR}%
  \BibitemOpen
  \bibfield{author}{%
  \bibinfo {author} {\bibfnamefont{W.~J.}\ \bibnamefont{Boettinger}}, \bibinfo
  {author} {\bibfnamefont{J.~A.}\ \bibnamefont{Warren}}, \bibinfo {author}
  {\bibfnamefont{C.}~\bibnamefont{Beckermann}},\ and\ \bibinfo {author}
  {\bibfnamefont{A.}~\bibnamefont{Karma}},\ }%
  \bibfield{journal}{%
  \Doi{10.1146/annurev.matsci.32.101901.155803}{\bibinfo {journal} {Annu. Rev.
  Mater. Res.}}\ }%
  \textbf{\bibinfo {volume} {32}},\ \bibinfo {pages} {163} (\bibinfo {month}
  {Aug}\ \bibinfo {year} {2002})%
  \bibAnnoteFile{NoStop}{Boettinger2002ARMR}%
\bibitem{Steinbach2009MSMSE}%
  \BibitemOpen
  \bibfield{author}{%
  \bibinfo {author} {\bibfnamefont{I.}~\bibnamefont{Steinbach}},\ }%
  \bibfield{journal}{%
  \Doi{10.1088/0965-0393/17/7/073001}{\bibinfo {journal} {Modell. Simul. Mater.
  Sci. Eng.}}\ }%
  \textbf{\bibinfo {volume} {17}},\ \bibinfo {pages} {073001} (\bibinfo {year}
  {2009})%
  \bibAnnoteFile{NoStop}{Steinbach2009MSMSE}%
\bibitem{Asbrink1999PRB}%
  \BibitemOpen
  \bibfield{author}{%
  \bibinfo {author} {\bibfnamefont{S.}~\bibnamefont{{\AA}sbrink}}, \bibinfo
  {author} {\bibfnamefont{A.}~\bibnamefont{Wa{\'s}kowska}}, \bibinfo {author}
  {\bibfnamefont{L.}~\bibnamefont{Gerward}}, \bibinfo {author}
  {\bibfnamefont{J.~S.}\ \bibnamefont{Olsen}},\ and\ \bibinfo {author}
  {\bibnamefont{Talik}},\ }%
  \bibfield{journal}{%
  \Doi{10.1103/PhysRevB.60.12651}{\bibinfo {journal} {Phys. Rev. B}}\ }%
  \textbf{\bibinfo {volume} {60}},\ \bibinfo {pages} {12651} (\bibinfo {year}
  {1999})%
  \bibAnnoteFile{NoStop}{Asbrink1999PRB}%
\bibitem{Kresse1993PRB}%
  \BibitemOpen
  \bibfield{author}{%
  \bibinfo {author} {\bibfnamefont{G.}~\bibnamefont{Kresse}}\ and\ \bibinfo
  {author} {\bibfnamefont{J.}~\bibnamefont{Hafner}},\ }%
  \bibfield{journal}{%
  \Doi{10.1103/PhysRevB.47.558}{\bibinfo {journal} {Phys. Rev. B}}\ }%
  \textbf{\bibinfo {volume} {47}},\ \bibinfo {pages} {558} (\bibinfo {year}
  {1993})%
  \bibAnnoteFile{NoStop}{Kresse1993PRB}%
\bibitem{Kresse1994PRB}%
  \BibitemOpen
  \bibfield{author}{%
  \bibinfo {author} {\bibfnamefont{G.}~\bibnamefont{Kresse}}\ and\ \bibinfo
  {author} {\bibfnamefont{J.}~\bibnamefont{Hafner}},\ }%
  \bibfield{journal}{%
  \Doi{10.1103/PhysRevB.49.14251}{\bibinfo {journal} {Phys. Rev. B}}\ }%
  \textbf{\bibinfo {volume} {49}},\ \bibinfo {pages} {14251} (\bibinfo {year}
  {1994})%
  \bibAnnoteFile{NoStop}{Kresse1994PRB}%
\bibitem{Kresse1996CMS}%
  \BibitemOpen
  \bibfield{author}{%
  \bibinfo {author} {\bibfnamefont{G.}~\bibnamefont{Kresse}}\ and\ \bibinfo
  {author} {\bibfnamefont{J.}~\bibnamefont{Furthm{\"u}ller}},\ }%
  \bibfield{journal}{%
  \Doi{10.1016/0927-0256(96)00008-0}{\bibinfo {journal} {Comp. Mater. Sci.}}\
  }%
  \textbf{\bibinfo {volume} {6}},\ \bibinfo {pages} {15} (\bibinfo {year}
  {1996})%
  \bibAnnoteFile{NoStop}{Kresse1996CMS}%
\bibitem{Kresse1996PRB}%
  \BibitemOpen
  \bibfield{author}{%
  \bibinfo {author} {\bibfnamefont{G.}~\bibnamefont{Kresse}}\ and\ \bibinfo
  {author} {\bibfnamefont{J.}~\bibnamefont{Furthm{\"u}ller}},\ }%
  \bibfield{journal}{%
  \Doi{PhysRevB.54.11169}{\bibinfo {journal} {Phys. Rev. B}}\ }%
  \textbf{\bibinfo {volume} {54}},\ \bibinfo {pages} {11169} (\bibinfo {year}
  {1996})%
  \bibAnnoteFile{NoStop}{Kresse1996PRB}%
\bibitem{Perdew1992PRB}%
  \BibitemOpen
  \bibfield{author}{%
  \bibinfo {author} {\bibfnamefont{J.~P.}\ \bibnamefont{Perdew}}, \bibinfo
  {author} {\bibfnamefont{J.}~\bibnamefont{Chevary}}, \bibinfo {author}
  {\bibfnamefont{S.}~\bibnamefont{Vosko}}, \bibinfo {author}
  {\bibfnamefont{K.~A.}\ \bibnamefont{Jackson}}, \bibinfo {author}
  {\bibfnamefont{M.~R.}\ \bibnamefont{Pederson}}, \bibinfo {author}
  {\bibfnamefont{D.}~\bibnamefont{Singh}},\ and\ \bibinfo {author}
  {\bibfnamefont{C.}~\bibnamefont{Fiolhais}},\ }%
  \bibfield{journal}{%
  \Doi{10.1103/PhysRevB.46.6671}{\bibinfo {journal} {Phys. Rev. B}}\ }%
  \textbf{\bibinfo {volume} {46}},\ \bibinfo {pages} {6671} (\bibinfo {year}
  {1992})%
  \bibAnnoteFile{NoStop}{Perdew1992PRB}%
\bibitem{Perdew1993PRB}%
  \BibitemOpen
  \bibfield{author}{%
  \bibinfo {author} {\bibfnamefont{J.~P.}\ \bibnamefont{Perdew}}, \bibinfo
  {author} {\bibfnamefont{J.}~\bibnamefont{Chevary}}, \bibinfo {author}
  {\bibfnamefont{S.}~\bibnamefont{Vosko}}, \bibinfo {author}
  {\bibfnamefont{K.~A.}\ \bibnamefont{Jackson}}, \bibinfo {author}
  {\bibfnamefont{M.~R.}\ \bibnamefont{Pederson}}, \bibinfo {author}
  {\bibfnamefont{D.}~\bibnamefont{Singh}},\ and\ \bibinfo {author}
  {\bibfnamefont{C.}~\bibnamefont{Fiolhais}},\ }%
  \bibfield{journal}{%
  \Doi{10.1103/PhysRevB.48.4978.2}{\bibinfo {journal} {Phys. Rev. B}}\ }%
  \textbf{\bibinfo {volume} {48}},\ \bibinfo {pages} {4978} (\bibinfo {year}
  {1993})%
  \bibAnnoteFile{NoStop}{Perdew1993PRB}%
\bibitem{Kresse1999PRB}%
  \BibitemOpen
  \bibfield{author}{%
  \bibinfo {author} {\bibfnamefont{G.}~\bibnamefont{Kresse}}\ and\ \bibinfo
  {author} {\bibfnamefont{D.}~\bibnamefont{Joubert}},\ }%
  \bibfield{journal}{%
  \Doi{10.1103/PhysRevB.59.1758}{\bibinfo {journal} {Phys. Rev. B}}\ }%
  \textbf{\bibinfo {volume} {59}},\ \bibinfo {pages} {1758} (\bibinfo {year}
  {1999})%
  \bibAnnoteFile{NoStop}{Kresse1999PRB}%
\bibitem{Choi2006PRB}%
  \BibitemOpen
  \bibfield{author}{%
  \bibinfo {author} {\bibfnamefont{H.}~\bibnamefont{Choi}}, \bibinfo {author}
  {\bibfnamefont{J.}~\bibnamefont{Shim}},\ and\ \bibinfo {author}
  {\bibfnamefont{B.}~\bibnamefont{Min}},\ }%
  \bibfield{journal}{%
  \Doi{10.1103/PhysRevB.74.172103}{\bibinfo {journal} {Phys. Rev. B}}\ }%
  \textbf{\bibinfo {volume} {74}},\ \bibinfo {pages} {172103} (\bibinfo {year}
  {2006})%
  \bibAnnoteFile{NoStop}{Choi2006PRB}%
\bibitem{Li2011MCP}%
  \BibitemOpen
  \bibfield{author}{%
  \bibinfo {author} {\bibfnamefont{H.}~\bibnamefont{Li}}, \bibinfo {author}
  {\bibfnamefont{B.}~\bibnamefont{Song}}, \bibinfo {author}
  {\bibfnamefont{W.}~\bibnamefont{Wang}},\ and\ \bibinfo {author}
  {\bibfnamefont{X.}~\bibnamefont{Chen}},\ }%
  \bibfield{journal}{%
  \Doi{10.1016/j.matchemphys.2011.04.072}{\bibinfo {journal} {Mater. Chem.
  Phys.}}\ }%
  \textbf{\bibinfo {volume} {130}},\ \bibinfo {pages} {39} (\bibinfo {month}
  {Oct}\ \bibinfo {year} {2011})%
  \bibAnnoteFile{NoStop}{Li2011MCP}%
\bibitem{Ruban2008RPP}%
  \BibitemOpen
  \bibfield{author}{%
  \bibinfo {author} {\bibfnamefont{A.~V.}\ \bibnamefont{Ruban}}\ and\ \bibinfo
  {author} {\bibfnamefont{I.}~\bibnamefont{Abrikosov}},\ }%
  \bibfield{journal}{%
  \Doi{10.1088/0034-4885/71/4/046501}{\bibinfo {journal} {Rep. Prog. Phys.}}\
  }%
  \textbf{\bibinfo {volume} {71}},\ \bibinfo {pages} {046501} (\bibinfo {year}
  {2008})%
  \bibAnnoteFile{NoStop}{Ruban2008RPP}%
\bibitem{Walle2002Calphad}%
  \BibitemOpen
  \bibfield{author}{%
  \bibinfo {author} {\bibfnamefont{A.}~\bibnamefont{van~de Walle}}, \bibinfo
  {author} {\bibfnamefont{M.}~\bibnamefont{Asta}},\ and\ \bibinfo {author}
  {\bibfnamefont{G.}~\bibnamefont{Ceder}},\ }%
  \bibfield{journal}{%
  \Doi{10.1016/s0364-5916(02)80006-2}{\bibinfo {journal} {Calphad}}\ }%
  \textbf{\bibinfo {volume} {26}},\ \bibinfo {pages} {539} (\bibinfo {month}
  {Dec}\ \bibinfo {year} {2002})%
  \bibAnnoteFile{NoStop}{Walle2002Calphad}%
\bibitem{Walle2002JPE}%
  \BibitemOpen
  \bibfield{author}{%
  \bibinfo {author} {\bibfnamefont{A.}~\bibnamefont{van~de Walle}}\ and\
  \bibinfo {author} {\bibfnamefont{G.}~\bibnamefont{Ceder}},\ }%
  \bibfield{journal}{%
  \Doi{10.1361/105497102770331596}{\bibinfo {journal} {J. Phase Equilib.}}\ }%
  \textbf{\bibinfo {volume} {23}},\ \bibinfo {pages} {348} (\bibinfo {year}
  {2002})%
  \bibAnnoteFile{NoStop}{Walle2002JPE}%
\bibitem{Walle2002MSMSE}%
  \BibitemOpen
  \bibfield{author}{%
  \bibinfo {author} {\bibfnamefont{A.}~\bibnamefont{Van~de Walle}}\ and\
  \bibinfo {author} {\bibfnamefont{M.}~\bibnamefont{Asta}},\ }%
  \bibfield{journal}{%
  \Doi{10.1088/0965-0393/10/5/304}{\bibinfo {journal} {Modell. Simul. Mater.
  Sci. Eng.}}\ }%
  \textbf{\bibinfo {volume} {10}},\ \bibinfo {pages} {521} (\bibinfo {year}
  {2002})%
  \bibAnnoteFile{NoStop}{Walle2002MSMSE}%
\bibitem{Walle2013JOM}%
  \BibitemOpen
  \bibfield{author}{%
  \bibinfo {author} {\bibfnamefont{A.}~\bibnamefont{van~de Walle}},\ }%
  \bibfield{journal}{%
  \Doi{10.1007/s11837-013-0764-3}{\bibinfo {journal} {JOM}}\ }%
  \textbf{\bibinfo {volume} {65}},\ \bibinfo {pages} {1523} (\bibinfo {year}
  {2013}),\ ISSN \bibinfo {issn} {1047-4838}%
  \bibAnnoteFile{NoStop}{Walle2013JOM}%
\bibitem{Fritsch2000SSI}%
  \BibitemOpen
  \bibfield{author}{%
  \bibinfo {author} {\bibfnamefont{S.}~\bibnamefont{Guillemet-Fritsch}},
  \bibinfo {author} {\bibfnamefont{C.}~\bibnamefont{Chanel}}, \bibinfo {author}
  {\bibfnamefont{J.}~\bibnamefont{Sarrias}}, \bibinfo {author}
  {\bibfnamefont{S.}~\bibnamefont{Bayonne}}, \bibinfo {author}
  {\bibfnamefont{A.}~\bibnamefont{Rousset}}, \bibinfo {author}
  {\bibfnamefont{X.}~\bibnamefont{Alcobe}},\ and\ \bibinfo {author}
  {\bibfnamefont{M.~M.}\ \bibnamefont{Sarriòn}},\ }%
  \bibfield{journal}{%
  \Doi{10.1016/S0167-2738(99)00340-9}{\bibinfo {journal} {Solid State Ionics}}\
  }%
  \textbf{\bibinfo {volume} {128}},\ \bibinfo {pages} {233 } (\bibinfo {year}
  {2000}),\ ISSN \bibinfo {issn} {0167-2738}%
  \bibAnnoteFile{NoStop}{Fritsch2000SSI}%
\bibitem{Errandonea2009PRB}%
  \BibitemOpen
  \bibfield{author}{%
  \bibinfo {author} {\bibfnamefont{D.}~\bibnamefont{Errandonea}}, \bibinfo
  {author} {\bibfnamefont{R.~S.}\ \bibnamefont{Kumar}}, \bibinfo {author}
  {\bibfnamefont{F.}~\bibnamefont{Manj{\'o}n}}, \bibinfo {author}
  {\bibfnamefont{V.}~\bibnamefont{Ursaki}},\ and\ \bibinfo {author}
  {\bibfnamefont{E.}~\bibnamefont{Rusu}},\ }%
  \bibfield{journal}{%
  \Doi{10.1103/PhysRevB.79.024103}{\bibinfo {journal} {Phys. Rev. B}}\ }%
  \textbf{\bibinfo {volume} {79}},\ \bibinfo {pages} {024103} (\bibinfo {year}
  {2009})%
  \bibAnnoteFile{NoStop}{Errandonea2009PRB}%
\bibitem{Kanamori1960JAP}%
  \BibitemOpen
  \bibfield{author}{%
  \bibinfo {author} {\bibfnamefont{J.}~\bibnamefont{Kanamori}},\ }%
  \bibfield{journal}{%
  \Doi{10.1063/1.1984590}{\bibinfo {journal} {J. Appl. Phys.}}\ }%
  \textbf{\bibinfo {volume} {31}},\ \bibinfo {pages} {S14} (\bibinfo {year}
  {1960})%
  \bibAnnoteFile{NoStop}{Kanamori1960JAP}%
\bibitem{Irani1962JPCS}%
  \BibitemOpen
  \bibfield{author}{%
  \bibinfo {author} {\bibfnamefont{K.}~\bibnamefont{Irani}}, \bibinfo {author}
  {\bibfnamefont{A.}~\bibnamefont{Sinha}},\ and\ \bibinfo {author}
  {\bibfnamefont{A.}~\bibnamefont{Biswas}},\ }%
  \bibfield{journal}{%
  \Doi{10.1016/0022-3697(62)90530-9}{\bibinfo {journal} {J. Phys. Chem.
  Solids}}\ }%
  \textbf{\bibinfo {volume} {23}},\ \bibinfo {pages} {711} (\bibinfo {month}
  {Jun}\ \bibinfo {year} {1962})%
  \bibAnnoteFile{NoStop}{Irani1962JPCS}%
\bibitem{Englman1972The}%
  \BibitemOpen
  \bibfield{author}{%
  \bibinfo {author} {\bibfnamefont{R.}~\bibnamefont{Englman}},\ }%
  \emph{\bibinfo {title} {The Jahn-Teller effect in molecules and crystals}}\
  (\bibinfo {publisher} {Wiley-Interscience New York},\ \bibinfo {year}
  {1972})%
  \bibAnnoteFile{NoStop}{Englman1972The}%
\bibitem{Gehring1975RPP}%
  \BibitemOpen
  \bibfield{author}{%
  \bibinfo {author} {\bibfnamefont{G.~A.}\ \bibnamefont{Gehring}}\ and\
  \bibinfo {author} {\bibfnamefont{K.~A.}\ \bibnamefont{Gehring}},\ }%
  \bibfield{journal}{%
  \Doi{10.1088/0034-4885/38/1/001}{\bibinfo {journal} {Rep. Prog. Phys.}}\ }%
  \textbf{\bibinfo {volume} {38}},\ \bibinfo {pages} {1} (\bibinfo {month}
  {Jan}\ \bibinfo {year} {1975})%
  \bibAnnoteFile{NoStop}{Gehring1975RPP}%
\bibitem{Kugel1982SPU}%
  \BibitemOpen
  \bibfield{author}{%
  \bibinfo {author} {\bibfnamefont{K.~I.}\ \bibnamefont{Kugel}}\ and\ \bibinfo
  {author} {\bibfnamefont{D.~I.}\ \bibnamefont{Khomskii}},\ }%
  \bibfield{journal}{%
  \Doi{10.1070/pu1982v025n04abeh004537}{\bibinfo {journal} {Sov. Phys. Usp.}}\
  }%
  \textbf{\bibinfo {volume} {25}},\ \bibinfo {pages} {231} (\bibinfo {month}
  {Apr}\ \bibinfo {year} {1982})%
  \bibAnnoteFile{NoStop}{Kugel1982SPU}%
\bibitem{Note1}%
  \BibitemOpen
  \bibinfo {note} {We would rather not blunt Occam's razor and trample
  intuition by suggesting a long-range-only phase-mixing effect with absolutely
  no evidence.}%
  \bibAnnoteFile{Stop}{Note1}%
\bibitem{Wojtowicz1959PR}%
  \BibitemOpen
  \bibfield{author}{%
  \bibinfo {author} {\bibfnamefont{P.}~\bibnamefont{Wojtowicz}},\ }%
  \bibfield{journal}{%
  \Doi{10.1103/physrev.116.32}{\bibinfo {journal} {Phys. Rev.}}\ }%
  \textbf{\bibinfo {volume} {116}},\ \bibinfo {pages} {32} (\bibinfo {month}
  {Oct}\ \bibinfo {year} {1959})%
  \bibAnnoteFile{NoStop}{Wojtowicz1959PR}%
\bibitem{Note2}%
  \BibitemOpen
  \bibinfo {note} {Although our calculations show zero-temperature distortion,
  while the empirical fitting is for room-temperature distortion, the same
  experiments show relatively little change in distortion below the transition
  temperature, so the equation is still valid.}%
  \bibAnnoteFile{Stop}{Note2}%
\bibitem{Englman1970PRB}%
  \BibitemOpen
  \bibfield{author}{%
  \bibinfo {author} {\bibfnamefont{R.}~\bibnamefont{Englman}}\ and\ \bibinfo
  {author} {\bibfnamefont{B.}~\bibnamefont{Halperin}},\ }%
  \bibfield{journal}{%
  \Doi{10.1103/physrevb.2.75}{\bibinfo {journal} {Phys. Rev. B}}\ }%
  \textbf{\bibinfo {volume} {2}},\ \bibinfo {pages} {75} (\bibinfo {month}
  {Jul}\ \bibinfo {year} {1970})%
  \bibAnnoteFile{NoStop}{Englman1970PRB}%
\bibitem{Noh2006APL}%
  \BibitemOpen
  \bibfield{author}{%
  \bibinfo {author} {\bibfnamefont{H.-J.}\ \bibnamefont{Noh}}, \bibinfo
  {author} {\bibfnamefont{S.}~\bibnamefont{Yeo}}, \bibinfo {author}
  {\bibfnamefont{J.-S.}\ \bibnamefont{Kang}}, \bibinfo {author}
  {\bibfnamefont{C.}~\bibnamefont{Zhang}}, \bibinfo {author}
  {\bibfnamefont{S.-W.}\ \bibnamefont{Cheong}}, \bibinfo {author}
  {\bibfnamefont{S.-J.}\ \bibnamefont{Oh}},\ and\ \bibinfo {author}
  {\bibfnamefont{P.}~\bibnamefont{Johnson}},\ }%
  \bibfield{journal}{%
  \Doi{10.1063/1.2178474}{\bibinfo {journal} {Appl. Phys. Lett.}}\ }%
  \textbf{\bibinfo {volume} {88}},\ \bibinfo {pages} {081911} (\bibinfo {year}
  {2006})%
  \bibAnnoteFile{NoStop}{Noh2006APL}%
\bibitem{Yeo2009JPCM}%
  \BibitemOpen
  \bibfield{author}{%
  \bibinfo {author} {\bibfnamefont{S.}~\bibnamefont{Yeo}}, \bibinfo {author}
  {\bibfnamefont{S.}~\bibnamefont{Guha}},\ and\ \bibinfo {author}
  {\bibfnamefont{S.}~\bibnamefont{Cheong}},\ }%
  \bibfield{journal}{%
  \Doi{10.1088/0953-8984/21/12/125402}{\bibinfo {journal} {J. Phys. Condens.
  Matter}}\ }%
  \textbf{\bibinfo {volume} {21}},\ \bibinfo {pages} {125402} (\bibinfo {year}
  {2009})%
  \bibAnnoteFile{NoStop}{Yeo2009JPCM}%
\bibitem{Liu2007PRL}%
  \BibitemOpen
  \bibfield{author}{%
  \bibinfo {author} {\bibfnamefont{J.~Z.}\ \bibnamefont{Liu}}, \bibinfo
  {author} {\bibfnamefont{G.}~\bibnamefont{Trimarchi}},\ and\ \bibinfo {author}
  {\bibfnamefont{A.}~\bibnamefont{Zunger}},\ }%
  \bibfield{journal}{%
  \Doi{10.1103/PhysRevLett.99.145501}{\bibinfo {journal} {Phys. Rev. Lett.}}\
  }%
  \textbf{\bibinfo {volume} {99}},\ \bibinfo {pages} {145501} (\bibinfo {month}
  {Oct}\ \bibinfo {year} {2007})%
  \bibAnnoteFile{NoStop}{Liu2007PRL}%
\bibitem{Laks1992PRB}%
  \BibitemOpen
  \bibfield{author}{%
  \bibinfo {author} {\bibfnamefont{D.~B.}\ \bibnamefont{Laks}}, \bibinfo
  {author} {\bibfnamefont{L.}~\bibnamefont{Ferreira}}, \bibinfo {author}
  {\bibfnamefont{S.}~\bibnamefont{Froyen}},\ and\ \bibinfo {author}
  {\bibfnamefont{A.}~\bibnamefont{Zunger}},\ }%
  \bibfield{journal}{%
  \Doi{10.1103/PhysRevB.46.12587}{\bibinfo {journal} {Phys. Rev. B}}\ }%
  \textbf{\bibinfo {volume} {46}},\ \bibinfo {pages} {12587} (\bibinfo {year}
  {1992})%
  \bibAnnoteFile{NoStop}{Laks1992PRB}%
\bibitem{Walle2014PRB}%
  \BibitemOpen
  \bibfield{author}{%
  \bibinfo {author} {\bibfnamefont{A.}~\bibnamefont{van~de Walle}}, \bibinfo
  {author} {\bibfnamefont{Q.}~\bibnamefont{Hong}}, \bibinfo {author}
  {\bibfnamefont{L.}~\bibnamefont{Miljacic}}, \bibinfo {author}
  {\bibfnamefont{C.~B.}\ \bibnamefont{Gopal}}, \bibinfo {author}
  {\bibfnamefont{S.}~\bibnamefont{Demers}}, \bibinfo {author}
  {\bibfnamefont{G.}~\bibnamefont{Pomrehn}}, \bibinfo {author}
  {\bibfnamefont{A.}~\bibnamefont{Kowalski}},\ and\ \bibinfo {author}
  {\bibfnamefont{P.}~\bibnamefont{Tiwary}},\ }%
  \bibfield{journal}{%
  \Doi{10.1103/physrevb.89.184101}{\bibinfo {journal} {Phys. Rev. B}}\ }%
  \textbf{\bibinfo {volume} {89}},\ \bibinfo {pages} {184101} (\bibinfo {month}
  {May}\ \bibinfo {year} {2014})%
  \bibAnnoteFile{NoStop}{Walle2014PRB}%
\bibitem{Honig2013NM}%
  \BibitemOpen
  \bibfield{author}{%
  \bibinfo {author} {\bibfnamefont{M.}~\bibnamefont{Honig}}, \bibinfo {author}
  {\bibfnamefont{J.~A.}\ \bibnamefont{Sulpizio}}, \bibinfo {author}
  {\bibfnamefont{J.}~\bibnamefont{Drori}}, \bibinfo {author}
  {\bibfnamefont{A.}~\bibnamefont{Joshua}}, \bibinfo {author}
  {\bibfnamefont{E.}~\bibnamefont{Zeldov}},\ and\ \bibinfo {author}
  {\bibfnamefont{S.}~\bibnamefont{Ilani}},\ }%
  \bibfield{journal}{%
  \Doi{10.1038/nmat3810}{\bibinfo {journal} {Nat. Mater.}}\ }%
  \textbf{\bibinfo {volume} {12}},\ \bibinfo {pages} {1112} (\bibinfo {year}
  {2013})%
  \bibAnnoteFile{NoStop}{Honig2013NM}%
\bibitem{Marianetti2001PRB}%
  \BibitemOpen
  \bibfield{author}{%
  \bibinfo {author} {\bibfnamefont{C.}~\bibnamefont{Marianetti}}, \bibinfo
  {author} {\bibfnamefont{D.}~\bibnamefont{Morgan}},\ and\ \bibinfo {author}
  {\bibfnamefont{G.}~\bibnamefont{Ceder}},\ }%
  \bibfield{journal}{%
  \bibinfo {journal} {Phys. Rev. B}\ }%
  \textbf{\bibinfo {volume} {63}} (\bibinfo {month} {May}\ \bibinfo {year}
  {2001}),\ \doi{\bibinfo {doi} {10.1103/PhysRevB.63.224304}}%
  \bibAnnoteFile{NoStop}{Marianetti2001PRB}%
\bibitem{Vleck1939JCP}%
  \BibitemOpen
  \bibfield{author}{%
  \bibinfo {author} {\bibfnamefont{J.~H.~V.}\ \bibnamefont{Vleck}},\ }%
  \bibfield{journal}{%
  \Doi{10.1063/1.1750327}{\bibinfo {journal} {J. Chem. Phys.}}\ }%
  \textbf{\bibinfo {volume} {7}},\ \bibinfo {pages} {72} (\bibinfo {year}
  {1939})%
  \bibAnnoteFile{NoStop}{Vleck1939JCP}%
\bibitem{Liu2008PRB}%
  \BibitemOpen
  \bibfield{author}{%
  \bibinfo {author} {\bibfnamefont{J.~Z.}\ \bibnamefont{Liu}}\ and\ \bibinfo
  {author} {\bibfnamefont{A.}~\bibnamefont{Zunger}},\ }%
  \bibfield{journal}{%
  \Doi{10.1103/PhysRevB.77.205201}{\bibinfo {journal} {Phys. Rev. B}}\ }%
  \textbf{\bibinfo {volume} {77}},\ \bibinfo {pages} {205201} (\bibinfo {month}
  {May}\ \bibinfo {year} {2008})%
  \bibAnnoteFile{NoStop}{Liu2008PRB}%
\bibitem{Holliger2011PRB}%
  \BibitemOpen
  \bibfield{author}{%
  \bibinfo {author} {\bibfnamefont{L.}~\bibnamefont{Holliger}}\ and\ \bibinfo
  {author} {\bibfnamefont{R.}~\bibnamefont{Besson}},\ }%
  \bibfield{journal}{%
  \Doi{10.1103/PhysRevB.83.174202}{\bibinfo {journal} {Phys. Rev. B}}\ }%
  \textbf{\bibinfo {volume} {83}},\ \bibinfo {pages} {174202} (\bibinfo {year}
  {2011})%
  \bibAnnoteFile{NoStop}{Holliger2011PRB}%
\bibitem{Nelson2013PRB035125}%
  \BibitemOpen
  \bibfield{author}{%
  \bibinfo {author} {\bibfnamefont{L.~J.}\ \bibnamefont{Nelson}}, \bibinfo
  {author} {\bibfnamefont{G.~L.}\ \bibnamefont{Hart}}, \bibinfo {author}
  {\bibfnamefont{F.}~\bibnamefont{Zhou}},\ and\ \bibinfo {author}
  {\bibfnamefont{V.}~\bibnamefont{Ozoli{\c{n}}{\v{s}}}},\ }%
  \bibfield{journal}{%
  \Doi{10.1103/PhysRevB.87.035125}{\bibinfo {journal} {Phys. Rev. B}}\ }%
  \textbf{\bibinfo {volume} {87}},\ \bibinfo {pages} {035125} (\bibinfo {year}
  {2013})%
  \bibAnnoteFile{NoStop}{Nelson2013PRB035125}%
\bibitem{Nelson2013PRB155105}%
  \BibitemOpen
  \bibfield{author}{%
  \bibinfo {author} {\bibfnamefont{L.~J.}\ \bibnamefont{Nelson}}, \bibinfo
  {author} {\bibfnamefont{V.}~\bibnamefont{Ozoli{\c{n}}{\v{s}}}}, \bibinfo
  {author} {\bibfnamefont{C.~S.}\ \bibnamefont{Reese}}, \bibinfo {author}
  {\bibfnamefont{F.}~\bibnamefont{Zhou}},\ and\ \bibinfo {author}
  {\bibfnamefont{G.~L.}\ \bibnamefont{Hart}},\ }%
  \bibfield{journal}{%
  \Doi{10.1103/PhysRevB.88.155105}{\bibinfo {journal} {Phys. Rev. B}}\ }%
  \textbf{\bibinfo {volume} {88}},\ \bibinfo {pages} {155105} (\bibinfo {year}
  {2013})%
  \bibAnnoteFile{NoStop}{Nelson2013PRB155105}%
\bibitem{Yamanaka2008AMin}%
  \BibitemOpen
  \bibfield{author}{%
  \bibinfo {author} {\bibfnamefont{T.}~\bibnamefont{Yamanaka}}, \bibinfo
  {author} {\bibfnamefont{A.}~\bibnamefont{Uchida}},\ and\ \bibinfo {author}
  {\bibfnamefont{Y.}~\bibnamefont{Nakamoto}},\ }%
  \bibfield{journal}{%
  \Doi{10.2138/am.2008.2934}{\bibinfo {journal} {Am. Mineral.}}\ }%
  \textbf{\bibinfo {volume} {93}},\ \bibinfo {pages} {1874} (\bibinfo {month}
  {Nov}\ \bibinfo {year} {2008})%
  \bibAnnoteFile{NoStop}{Yamanaka2008AMin}%
\end{thebibliography}%

\end{document}